\documentclass[reprint,superscriptaddress,amsmath,amssymb,aps,
floatfix]
{revtex4-1}

\usepackage{graphicx}
\usepackage{dcolumn}
\usepackage{bm}
\usepackage{xcolor}
\usepackage{ulem}

\begin{document}

\preprint{APS/123-QED}

\title{Spatial distribution of Dust Density Wave Properties in Fluid Complex Plasmas}

\author{P. Bajaj}\email{prapti.bajaj@dlr.de}\affiliation{Institut f\"{u}r Materialphysik im Weltraum, Deutsches Zentrum f\"{u}r Luft- und Raumfahrt (DLR), D-82234 We\ss ling, Germany}
\author{S. Khrapak}
\affiliation{Joint Institute for High Temperatures, Russian Academy of Sciences, 125412 Moscow, Russia}
\author{V. Yaroshenko}\affiliation{Institut f\"{u}r Materialphysik im Weltraum, Deutsches Zentrum f\"{u}r Luft- und Raumfahrt (DLR), D-82234 We\ss ling, Germany}
\author{M. Schwabe}
\affiliation{Institut f\"{u}r Materialphysik im Weltraum, Deutsches Zentrum f\"{u}r Luft- und Raumfahrt (DLR), D-82234 We\ss ling, Germany}

\date{\today}

\begin{abstract}
Complex plasmas consist of microparticles embedded in a low-temperature plasma and allow investigating various effects by tracing the motion of these microparticles. Dust density waves appear in complex plasmas as self-excited acoustic waves in the microparticle fluid at low neutral gas pressures. Here we show that various properties of these waves depend on the position of the microparticle cloud with respect to the plasma sheath and explain this finding in terms of the underlying ion-drift instability. These results may be helpful in better understanding the propagation of dust density waves in complex plasmas and beyond, for instance, in astrophysical dusty plasmas.
\end{abstract}
\maketitle

\section{\label{sec:intro}Introduction}
Complex (or dusty) plasmas are formed by introducing micrometer-sized solid particles in the plasma environment of ions, electrons and neutrals. These microparticles attain a negative or positive charge depending on the charging mechanism \cite{Fortov2005}. Under typical laboratory conditions, they are strongly negatively charged due to the higher mobility of electrons. Owing to the high charge on the microparticles, they interact with one another and exhibit interesting collective behavior. Dusty plasmas are naturally found in astrophysical situations such as planetary rings, comet tails, and interplanetary and interstellar clouds \cite{Goertz1989, Goertz1983, Ivlev2018,Verheest2000,Bliokh1995}. These are also investigated in laboratories for materials processing, etching, fusion and other such experiments with plasma and dust \cite{Kersten2010, Winter1998}. To differentiate specially designed dusty plasmas, the term complex plasmas is used \cite{Fortov2005}. Owing to their low charge-to-mass ratio and relatively large size, the microparticle motion can be traced using optical high-speed cameras, thus enabling the study of solid, liquid or gaseous complex plasma systems at the atomistic (kinetic) level.

In complex plasmas, the ion-dust streaming instability \footnote{The ion-dust streaming instability or two-stream instability caused by the positively charged ions flowing towards the electrode excites and sustains DDWs.} causes the self-excitation of longitudinal waves called dust acoustic waves (DAWs) \cite{Barkan1995, Merlino2014,  Rosenberg1993, Rosenberg1997} or dust density waves (DDWs) \cite{Fortov2000, Rosenberg2014, Rosenberg1996}.
DAWs were first predicted by Rao, Shukla and Yu \cite{Rao1990} and have since been extensively studied theoretically as well as experimentally using ground based setups \cite{Schwabe2007, Williams2013, Williams2019, Tsai2020},
and microgravity experiments on the International Space Station (ISS) \cite{Khrapak2003, Schwabe2008}, and on parabolic flights \cite{Himpel2014}. The DDWs usually move in the direction of the ion flow. Interestingly, the presence of DDWs has also been linked to the formation of agglomerates \cite{Du2010}.

\begin{figure}
	\includegraphics[width=0.45\textwidth]{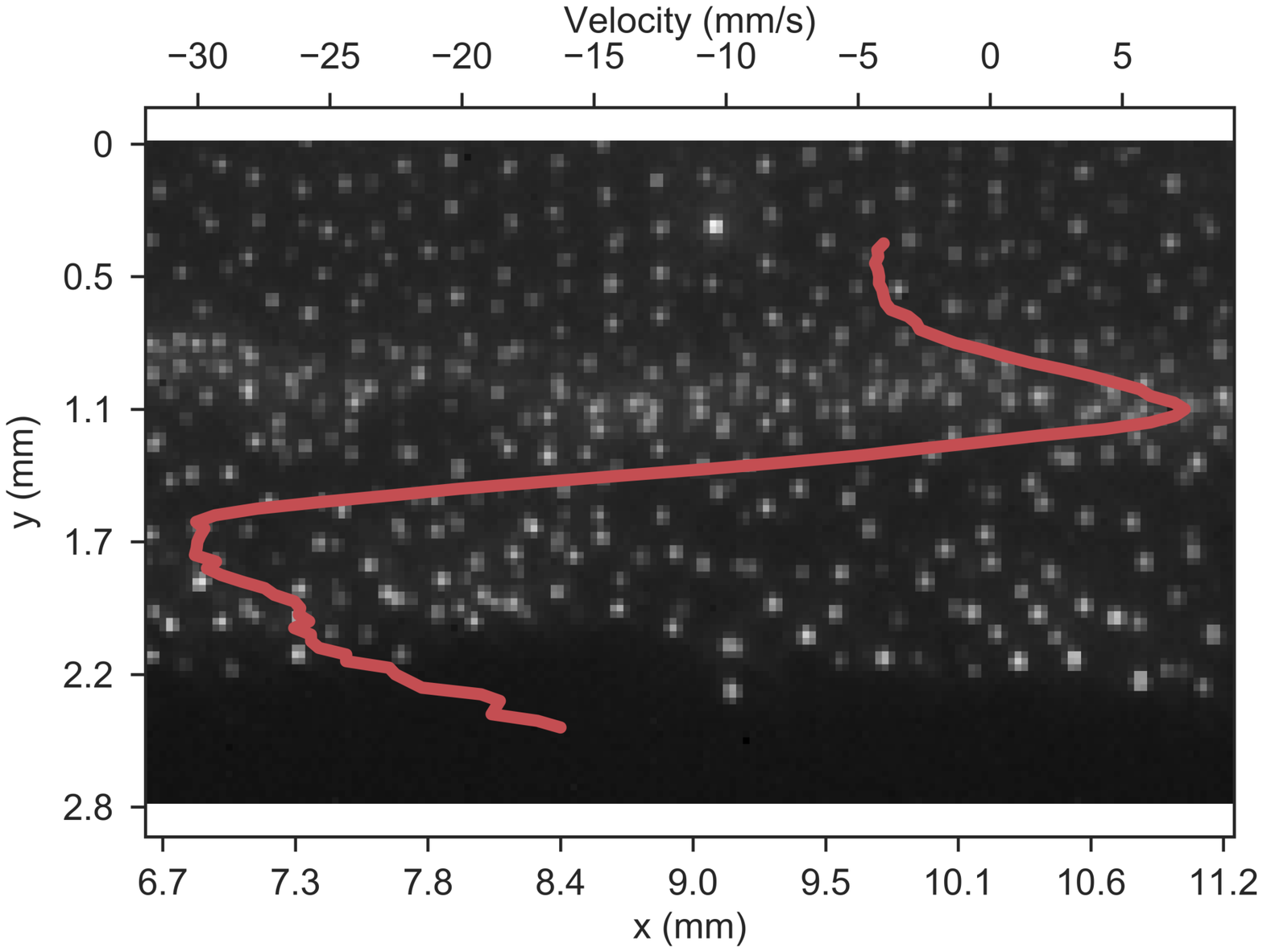}
	\caption{\label{fig:raw1} Raw image (cropped for clarity) from the high-speed camera at\textit{ $\Delta$T }= 0 K. The superimposed line shows the average particle velocity in the vertical (y) direction. As discussed in Section~\ref{sec:ana_meth}, this property was used to find the wave crests from raw images.
    }
\end{figure}

 Although there have been many studies on the relation between DDWs and the electric field that produces them, both theoretically \cite{Khrapak2020, Rosenberg1993} and experimentally \cite{Merlino1998, Jaiswal2018, Schwabe2007}, there is not much knowledge on the spatial distribution of the DDW properties \cite{Flanagan2010, Tadsen2017, Fortov2003, Thomas2006}. In \cite{Thomas2006}, Thomas et al. study the spatial growth of DDWs in a direct current (DC) glow discharge by varying the pressure of the neutral gas. It is interesting to note that the values of DDW speed obtained in their experiment are in the range of the values found using our experiment, which was conducted using different temperature gradients.
Here, we endeavor to understand the DDW properties as the microparticle cloud is moved from the sheath region to the bulk plasma region.

For the case of radio-frequency (RF) complex plasmas under gravity conditions, these waves can be observed in the pre-sheath or sheath region, the positive space charge region near the walls of the chamber, where the downward force of gravity is balanced by the upward electric force on the microparticles. Consequently, in order to understand these waves better, it is essential to study the effect of the sheath region on the DDWs.

In this article, we report a study where the microparticle cloud is lifted out of the sheath without changing any other relevant parameters which makes it possible to study the change in the DDW properties as a function of the microparticle cloud position. Theoretical considerations supplement the experimental observations.

An image of the microparticle cloud with DDW propagating towards the sheath, i.e., in the direction of ion flux is shown in Fig.~\ref{fig:raw1}. The wave crests are well visible as regions of higher microparticle density, separated by wave troughs with a lower particle density. The average particle velocity obtained from tracing the particle motion is superimposed on the raw image. As we shall see in the following, this allows to determine the position of the wave crests with good accuracy.

In Sec.~\ref{sec:setup}, we briefly describe the experimental setup and the experimental methods. This is followed by the detailed description of analytical techniques in Sec.~\ref{sec:ana_meth} which are used to arrive at the experimental and theoretical results presented in Sec.~\ref{sec:res}. We give a conclusion in Sec.~\ref{sec:conclusion}.

\section{\label{sec:meth}Experiment}
\subsection{\label{sec:setup}Setup}

\begin{figure}
	\includegraphics[width=0.45\textwidth]{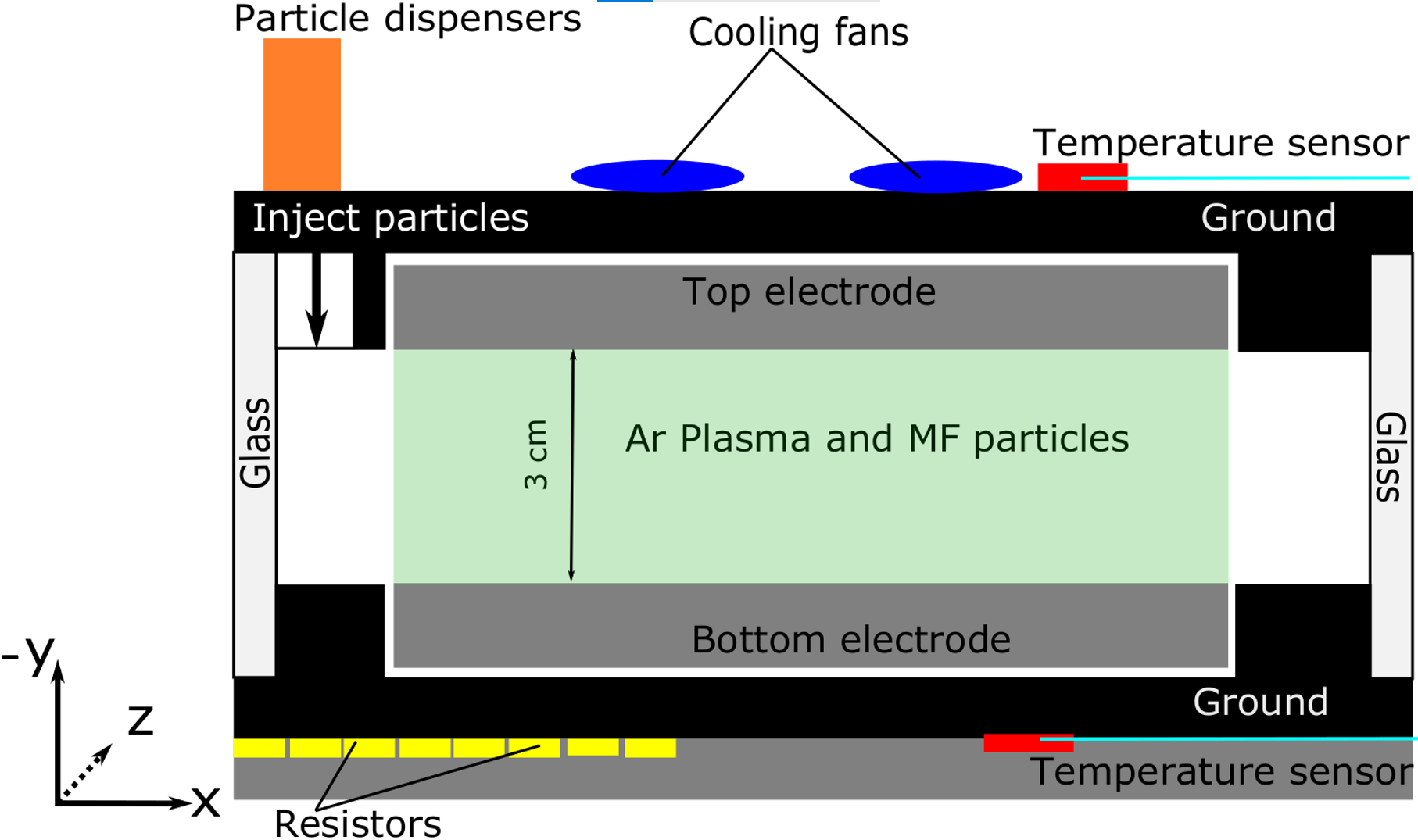}
	\caption{\label{fig:exp_setup} PK-3 Plus ground setup. This is the view from the direction of the high-speed camera. The eight resistors heat up the lower electrode, and the fans cool down the upper electrode, thus creating a temperature difference (T ). The light from a laser (not shown) is split into a vertical laser sheet (in green or light-gray) that is reflected by the particles and recorded with the high-speed camera. Gravity points in the positive y-direction, i.e., towards the electrode with embedded resistors.}
\end{figure}

The experimental setup of the ground-based PK-3 Plus chamber \cite{Thomas2008} is shown schematically in Fig.~\ref{fig:exp_setup}. A plasma was created in a capacitively-coupled RF chamber. The two electrodes were driven by a sinusoidal signal of 13.56~MHz, leading to a voltage of 56~V$_{pp}$ on the electrodes. We used argon at a pressure of 23~Pa with a flow rate of 0.2~sccm (standard cubic centimeters per minute), and injected mono-disperse melamine formaldehyde (MF) spheres of mass density 1510~kg/m$^3$ and diameter 4.4~$\mu$m. A laser with a wavelength of 686~nm and a power of 45~mW illuminated the microparticles in a vertical plane. Their motion was tracked using the CMOS video camera Photron Fastcam-1024 PCI at a speed of 2000 frames per second (fps) with a field of view of 640 x 640~pixels$^2$ at a spatial resolution of 28~$\mu$m/pixel. This high-speed imaging makes it possible to study the particle behavior at the kinetic level and track particles individually from frame to frame even inside the high-speed waves studied in this work.

This experimental setup is well suited to study DDWs due to its unique ability to change the temperature difference between the top and bottom electrodes. This is possible by heating the bottom electrode with eight resistors (see Fig.~\ref{fig:exp_setup}) and cooling the top electrode with two fans. The temperature difference is measured by the temperature sensors attached to the top and bottom ground plates \footnote{The temperature sensors are attached to the top and bottom ground plates and not the top and bottom electrodes. This leads to an error of up to 1~K in the measured values of temperature difference.}. The particles experience a thermophoretic force pointing upwards, against the force of gravity due to this temperature gradient \cite{Rothermel2002}. This causes the microparticle cloud to move away from the bottom electrode into the bulk plasma. The magnitude of the thermophoretic force is proportional to the temperature gradient $\nabla$T, square of the particle radius $r_p$, and inversely proportional to the gas kinetic cross section $\sigma$:

\begin{equation}
F_{\rm th} \approx -{\mathcal A}\frac{r_{p}^{2}}{\sigma}\nabla T,
\label{eq:thermo_force}
\end{equation}
where
\textit{$\nabla$T = {$\Delta$T}/{\text{(distance between electrodes)}}}. In the current experiment, the distance between the electrodes is 3~cm (as shown in Fig. \ref{fig:exp_setup}). The temperature gradient is expressed in energy units (if units K/cm are used, the front factor should also contain the Boltzmann constant $k_{\rm B}$). The numerical coefficient ${\mathcal A}$ may depend on details of the interaction between the neutral atoms and the particle surface. However, even for the simplest case of specular reflection, different values have been reported in the literature (for instance ${\mathcal A}=3.33$ in Ref.~\cite{Rothermel2002} and ${\mathcal A}=1.8$ in Refs.~\cite{Fortov2005,Fortov2004}). Here we used ${\mathcal A}\approx 2.8$ based on the Waldman's expression for the thermophoretic force~\cite{Waldmann1959} combined with the dilute hard-sphere gas limit of the thermal conductivity coefficient, appropriate for a mono-atomic gas~\cite{Lifshitz1995}. This value is also consistent with experimental measurements reported in Ref.~\cite{Rothermel2002}. For the gas kinetic cross section in argon we used $\sigma = 4.02\times 10^{-15}$ cm$^2$, as recommended in Ref.~\cite{Raizer2011}.

In this work, we took advantage of gravity to see the highest speed of DDWs when they were located just above the sheath region (at \textit{$\Delta$T} = 0~K). Then we increased the temperature difference until the particles reached the bulk region where the DDWs disappeared, and the microparticles formed a crystal-like structure.

\begin{figure}
	\includegraphics[width=0.45\textwidth]{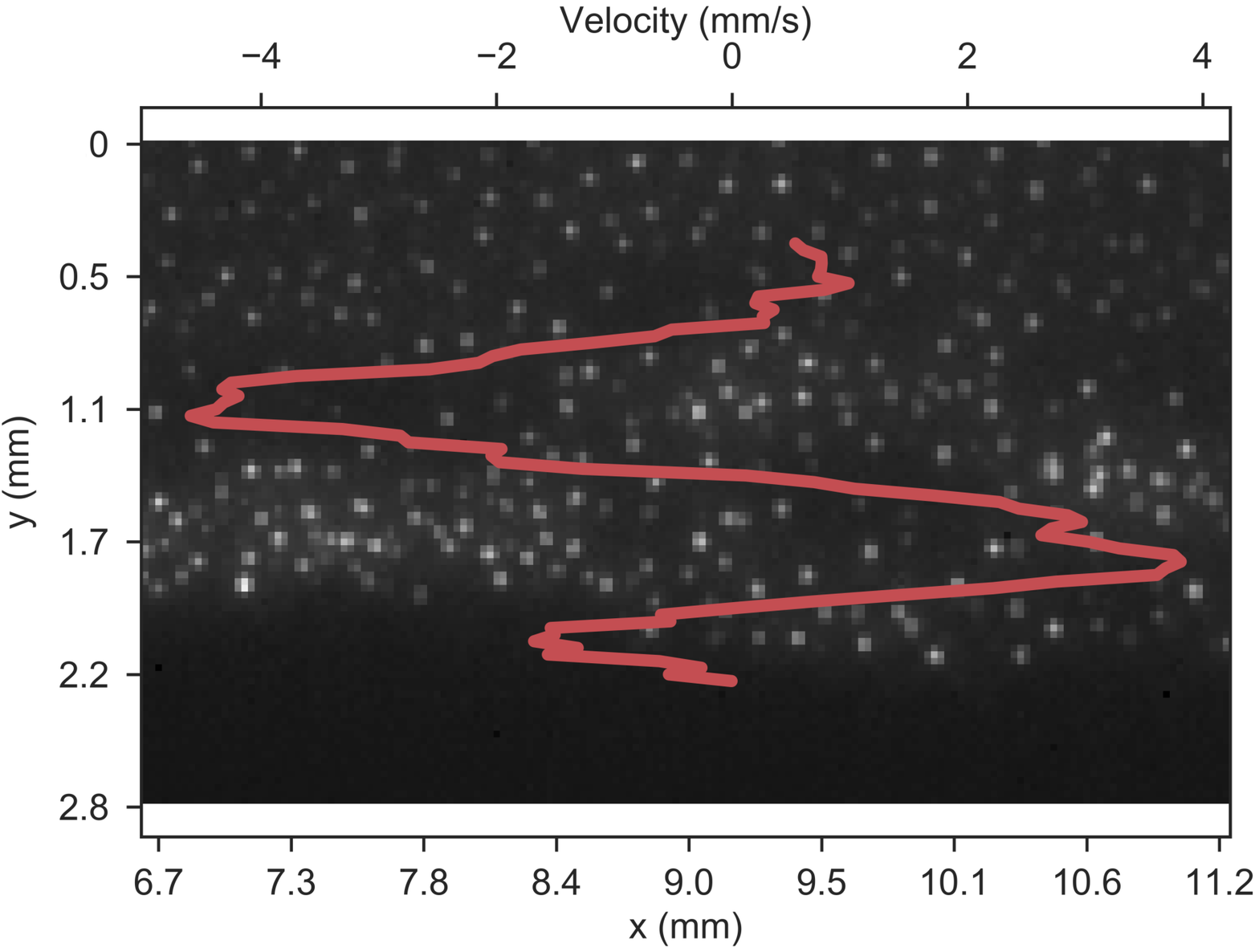}
	\caption{\label{fig:raw2} Raw image (cropped for clarity) from the high-speed camera at \textit{$\Delta$T} = 22.5~K. The line shows the average particle $y$-velocity. As discussed in Section~\ref{sec:ana_meth}, this method was used to find the wave crests. It became less efficient for higher temperature differences, compare with Figure~\ref{fig:raw1}.}
\end{figure}

To perform this experiment, after injecting particles, we reduced the pressure to 23~Pa at which point we started observing unstable waves in the microparticle cloud with no temperature difference between the electrodes (as shown in Fig.~\ref{fig:raw1}). After this, we changed the temperature difference (\textit{$\Delta$T}) in steps of 1 - 1.5~K, and recorded high-speed videos of the microparticle motion at each step. The temperature difference was increased until the DDWs disappeared, and we saw crystallization of the microparticles. The microparticle cloud and the superimposed average velocity in vertical direction at \\\textit{$\Delta$T} = 22.5~K can be seen in Fig.~\ref{fig:raw2}. In comparison with Fig.~\ref{fig:raw1}, it is clear that the wave crests were less well defined at higher temperature differences, when the waves were closer to disappearing, and thus it became more difficult to trace the wave motion, as we shall see in the following. It is also interesting to observe that the average particle velocity changes from negative values to positive values as the DDW travels downwards from y=0~mm. This is due to the motion of particles inside the wave compression and rarefaction as the DDWs travel towards the lower electrode. Negative velocities imply upward motion of particles whereas positive velocities imply downward motion of the particles.

We also performed another experiment at the same experimental parameters, but using only a few layers of microparticles compared to an extended cloud in the first experiment. This made it possible to calibrate the equilibrium vertical position of the microparticle cloud at the selected temperature gradients.

\subsection{\label{sec:ana_meth}Analytical and Imaging methods}

\begin{figure}
	\includegraphics[width=0.45\textwidth]{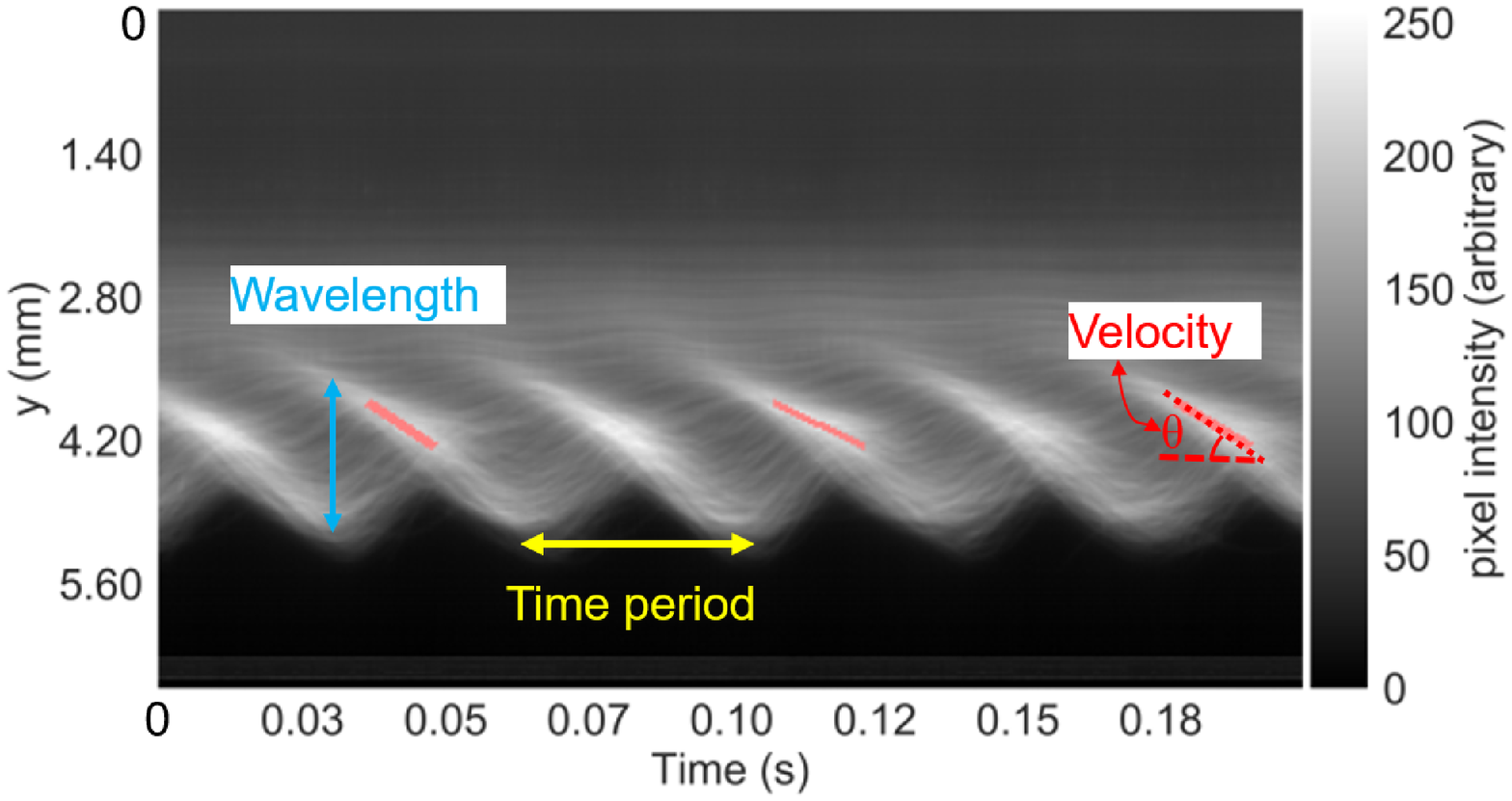}
	\caption{\label{fig:pdg} Periodgram or space-time plot for $\Delta$T = 2 K. The superimposed red lines show some of the slopes found from the process of template matching and were used to determine the wave speed for each temperature difference. Here, only 3 slopes are superimposed out of 5 found by template matching for clarity. The ways to measure wavelength, time period of the waves, and wave velocity are indicated in the image.}
\end{figure}

We used two complimentary methods to measure the speed of the DDWs: periodgrams and particle tracing, for each selected temperature gradient. As shown in Fig.~\ref{fig:pdg}, periodgrams or space-time plots were created by plotting the variation of average pixel intensity with time. To obtain this image, we averaged the pixel intensity over the entire horizontal direction at each $y$-coordinate of the raw image, e.g. Fig.~\ref{fig:raw1} and Fig.~\ref{fig:raw2}, for each image in the entire time series. We then plotted the change in average intensity as a function of time and height for each selected temperature gradient. In Fig.~\ref{fig:pdg}, we present the periodgram at $\Delta$T = 2~K. In this figure, the white regions (or high intensity regions) correspond to high density of particles, that is, the wave crests \cite{Schwabe2008}. Then, the positions of the wave crests could be used to calculate the various wave parameters like frequency, wave speed and wavelength. The distance between two wave crests in time (space) corresponds to the time period (wavelength) of the wave. The slope of these wave crests in the periodgram corresponds to the wave speed. Periodgrams for other temperature differences have been presented in the Supplemental Material \footnote{See Supplemental Material at [URL will be inserted by publisher] for periodgrams at other temperature differences.}.

\begin{figure}
	\includegraphics[width=0.45\textwidth]{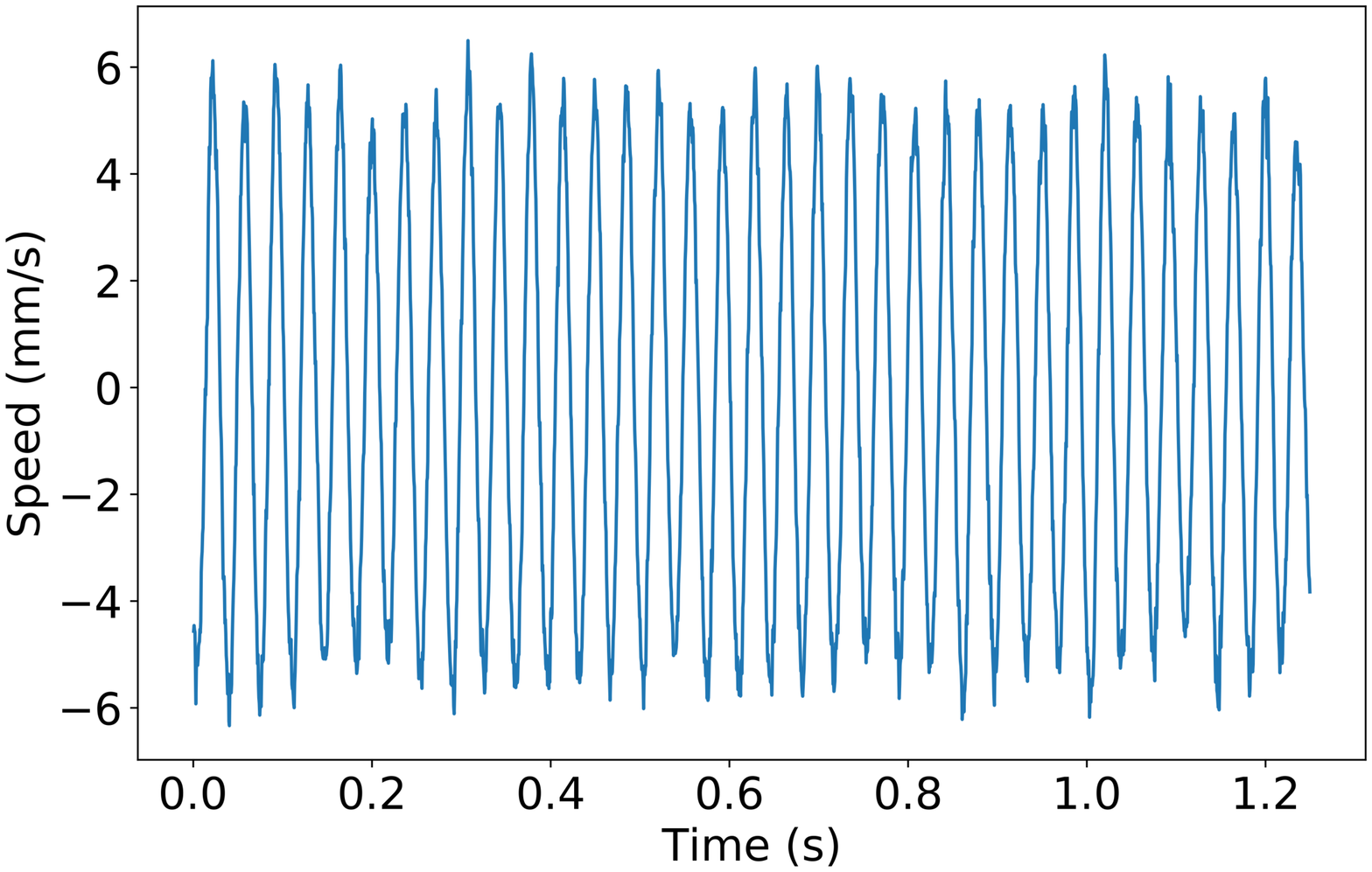}
	\caption{\label{fig:freq}Average particle speed resulting from particle tracking in the region 0.28~mm $\leq  y \leq $ 0.56~mm, as a function of time at $\Delta$T = 0~K.  }
\end{figure}

We cropped each periodgram to a region of interest of 0.7~mm height, where the wave crests are visible clearly, and then applied Gaussian blurring with a kernel size of $3 \times 3 \, \text{pixels}^2$ to these space-cropped periodgrams. Gaussian blurring dissolves any minute artifacts in the periodgrams in order to make the wave crests more recognizable. To perform template matching on these periodgrams, we first created only a single template of a one-pixel line in a box of $25 \times 25 \, \text{pixels}^2$ and performed Gaussian blurring with a kernel size of $5 \times 5 \, \text{pixels}^2$. This line was constructed at an angle of 45$^{\circ}$ with the negative x-axis for this particular template. We rotated this 45$^{\circ}$ angled line in this template and created eighty-nine more templates, each with a different angle varying from 5$^{\circ}$ to 50$^{\circ}$, corresponding to detectable velocities between 4.9~mm/s and 67~mm/s. Using these ninety templates with the package \textit{Multi-Template-Matching} \cite{Thomas2020} (MTM), we calculated the DDW phase velocity and frequency (as shown later in Figs. \ref{fig:results1} and \ref{fig:results2}).\\

The MTM package tests each template on each image-patch of the same size as the template to find the best match for the chosen image-patch. We scan the entire periodgram in such a manner by making image-patches at each pixel of the periodgram. After this, we obtain the correlation map (or matrix) of the templates matched at each pixel of the periodgram. This correlation map gives us the correlation score for each pixel value where the correlation was calculated between each template and the chosen image patch. Using this correlation map, we choose the pixel positions from the correlation scores higher than a threshold value to obtain all wave crests in the periodgram. The name of the template with the highest correlation score at a particular positon of the wave crest gives us the slope of the wave crest at that position. This is then used to calculate the wave velocity using all the wave crests in the periodgram. Using the time-coordinates of the wave crests, we can also calculate the wave frequency with this method. In order to find more in-depth details on this package and the procedure followed here, please refer to Ref. \footnote{Multi-template matching: How does it work? -  \protect\url{https://multi-template-matching.github.io/Multi-Template-Matching/doc/explanations}}.\\

Some of the slopes found by template matching are shown as red superimposed lines in Fig.~\ref{fig:pdg}.  Using the degree of correlation between the template and the matched wave crest as weight, we calculated the average speed of the wave crests at each selected temperature gradient. After matching the periodgram with the templates, we also obtained the space-time coordinates of the wave crests. These subsequent time-coordinates of the wave crests were used to calculate the frequency of the DDWs. However, as we found a more accurate method to measure the DDW frequency, which we describe later, we do not present our results of frequency measurements using template matching on periodgrams (more on this in the supplementary material).

In order to develop a deeper understanding of our results, we used another analytical method. Here, we individually tracked particles using the package \textit{trackpy} \cite{Allan2016}. Thereafter, we calculated each particle's velocity from one frame to the other. The efficiency of the particle tracking can be seen in Fig.~\ref{fig:freq} where average vertical particle velocity in a region of 0.28~mm height is plotted as a function of time. Here, we can see the average motion of particles moving inside the wave crests and troughs with time. We obtained the frequency of the DDWs by Fourier transforming these particle velocity data of the entire time sequence recorded for a specific \textit{$\Delta$T}.

We then used the distribution of vertical particle velocities in space to find the wave crests in the raw images. We defined the $y$-position of the wave crest as the position of the peak of positive particle velocities and following the movement of this peak downwards as the wave propagated to the bottom edge of the plasma. One such tracked wave crest is shown in Fig.~\ref{fig:raw1} as superimposed on the raw image with a second x-axis showing the magnitude of the average velocity in space. It is easy to observe here that the identification of the position of the wave crest worked satisfactorily as it is located exactly at the $y$-coordinate with the positive peak value of average velocity. However, this process became less efficient when the wave crest became irregular in space due to non-uniform particle density, as is visible, for example, in Fig.~\ref{fig:raw2} which shows the raw image at \textit{$\Delta$T} = 22.5~K. We saw this irregularity in particle density inside the wave crests at higher \textit{$\Delta$T} because the wave motion became more non-uniform as the thermophoretic force increased and the waves were excited less strongly. A non-uniform distribution of particle density inside a wave crest leads to a less well-defined velocity peak which in turn makes it difficult to track the wave crest position at higher temperature gradients. Consequently, the results presented in the next section using these tracking algorithms show a higher standard deviation for data points at higher temperature gradients (for example, in Fig.~\ref{fig:results1}). Finally, we performed linear fits on the positions of each wave crest with time to obtain the wave speed and calculated the average speed of all wave crests recorded for a given temperature gradient.

After obtaining the evolution of wave speed and frequency with temperature gradient, we measured the equilibrium position of the microparticle cloud with changing temperature gradients by using only a few (3 to 4) microparticle layers in the chamber. Here as well, we used \textit{trackpy} \cite{Allan2016} to locate the particles and find the vertical position of the microparticle cloud with changing temperature gradients. The resulting average levitation heights are shown in Figs.~\ref{fig:results1} and \ref{fig:results2}.

\section{\label{sec:res}Results}

\subsection{Experiment}

\begin{figure}
	\includegraphics[width=0.45\textwidth]{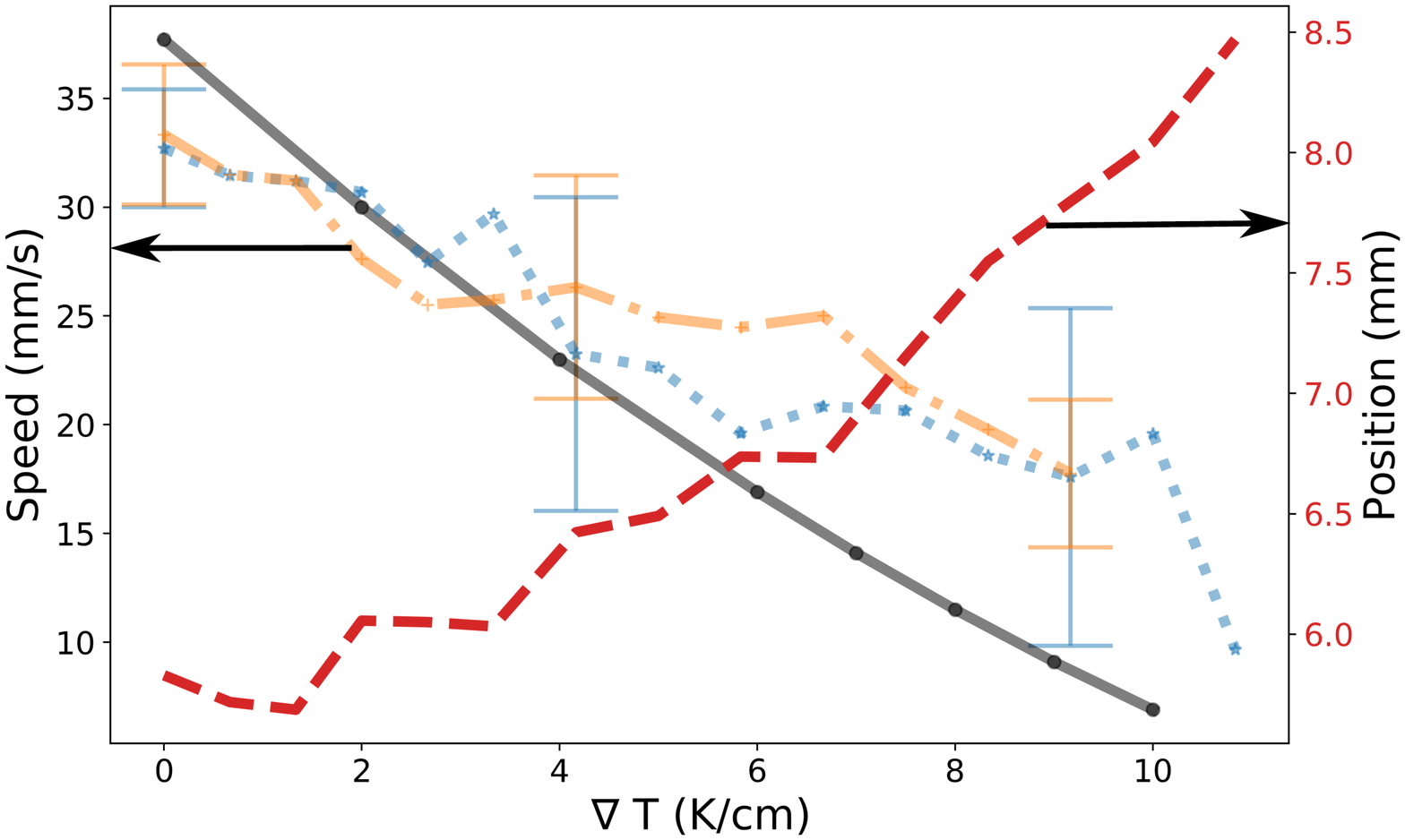}
	\caption{\label{fig:results1} Wave speed measured with velocity peak tracking (orange dash-dotted line) and template matching on the periodgram (blue dotted line), as well as the cloud position (red dashed line) with respect to the temperature gradient. It can be seen that the two methods of determining the wave speed agree with each other, and that the speed decreases with an increase in the levitation height. The vertical barred lines show the standard deviation for some selected temperature gradients. The black line shows the theoretical prediction obtained from Eq.~(\ref{sound}) as discussed in the text. The arrows point at the respective axes for the line plots.}
\end{figure}

Figure~\ref{fig:results1} shows the wave speed (in blue and orange) and cloud position (in red) as a function of temperature gradient. The wave speed decreased as the DDWs moved away from the sheath and into the bulk plasma region. In Fig.~\ref{fig:results1}, one can observe that the standard deviation (vertical bars) from template matching is larger than that for wave crest tracking. This is because of the increase in wave irregularity with increasing temperature difference which makes it more and more difficult to match artificially created templates with the wave crests in periodgrams.

\begin{figure}
	\includegraphics[width=0.45\textwidth]{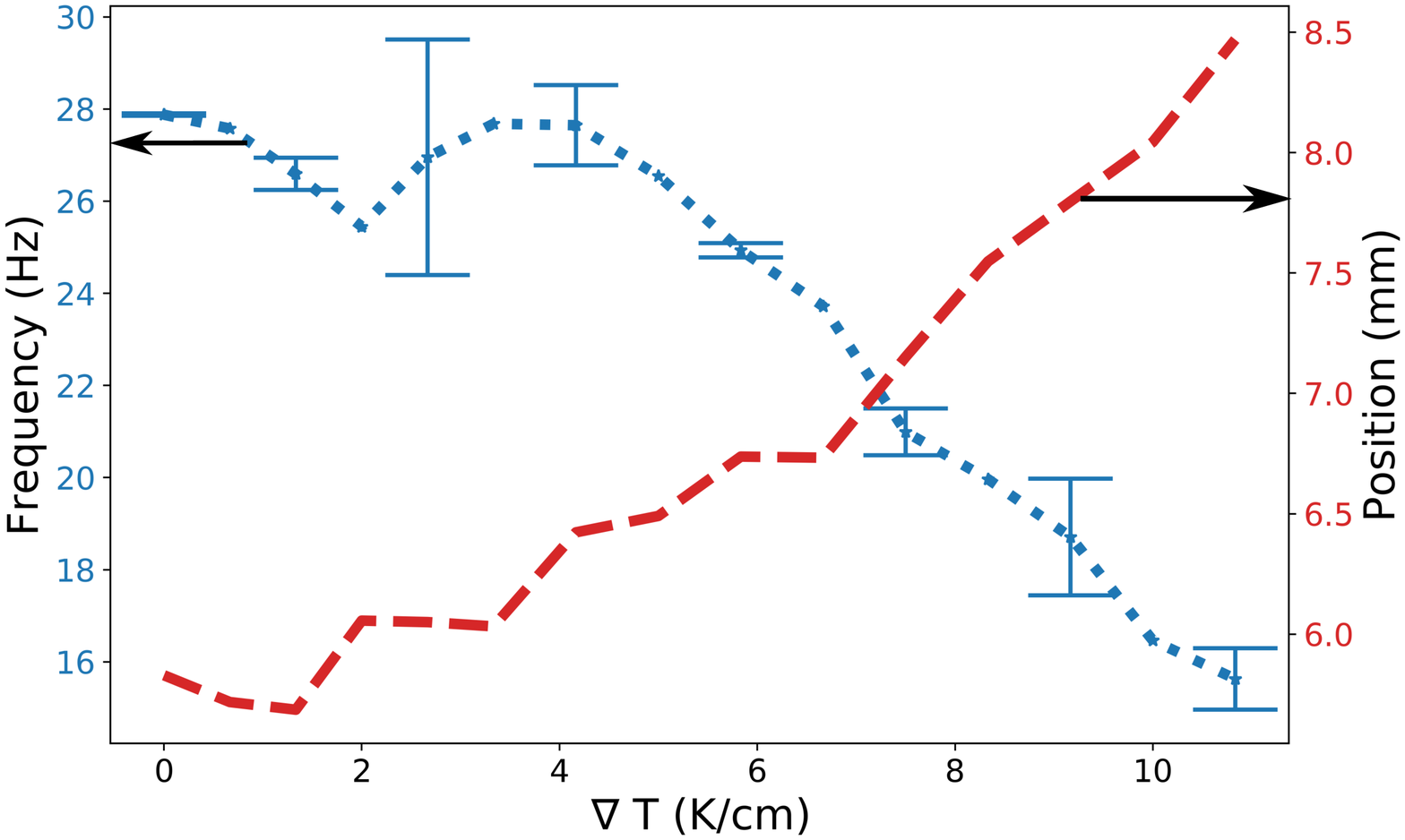}
	\caption{\label{fig:results2} Wave frequency (blue dotted line) and cloud position above the bottom electrode (red dashed line) with respect to temperature gradient. The wave frequency falls as the levitation height increases. The vertical barred lines show the standard deviation for some selected temperature gradients. The arrows point to the respective axes for the line plots.
    }
\end{figure}

Figure \ref{fig:results2} shows the dependence of wave frequency and cloud position on the temperature gradient. The wave frequency decreased as the DDWs moved away from the sheath and into the bulk plasma region. It is also interesting to note that the DDWs became more non-linear as they traveled towards the sheath. However, the non-linear properties of these DDWs will be studied in a future publication.

\subsection{Theory}

There is a relatively wide parameter regime at moderate neutral gas pressures where the ion and the particle components are dominated by collisions with neutrals, but the ion-dust streaming instability can nevertheless operate. This regime was considered in detail previously~\cite{Khrapak2003,Khrapak2020}, and we summarize only briefly the main results here.

The dispersion relation is obtained from
\begin{equation}\label{permit}
\epsilon(k,\omega)=1+\sum_j \chi_j\simeq \sum_j \chi_j=0,
\end{equation}
where $\omega$ is the wave frequency, $k$ is the wave number, the summation is over all charged species $j$, and the long-wavelength limit is considered (so that the quasineutrality condition holds and the Laplacian in the Poisson equation can be neglected).  For light mobile electrons it is sufficient to keep the static contribution, which is simply $\chi_e=1/{k^2\lambda_e^2}$, where $\lambda_e=\sqrt{T_e/4\pi e^2 n_e}$ is the electron Debye radius ($n_e$ and $T_e$ are the electron density and temperature). The latter can be normally neglected under the condition $T_e\gg T_i$, where $T_i$ is the ion temperature. For the ions we use the collisional response function~\cite{Khrapak2020}
\begin{equation}\label{ionresp}
\chi_i=\frac{k^{-2}\lambda_i^{-2}}{1-M^2+iM(\nu_i/kv_{{\rm T}i})},
\end{equation}
where $\lambda_i=\sqrt{T_i/4\pi e^2 n_i}$ is the ion Debye radius,  $n_i$ is the ion density, $M=u_i/v_{{\rm T}i}$ is the ion {\it thermal} Mach number ($u_i$ and $v_{{\rm T}i}=\sqrt{T_i/m_i}$ are the ion drift and thermal velocities, respectively), and $\nu_i$ is the effective ion-neutral collision frequency. For the particle component, neglecting pressure term in the momentum equation, yields
\begin{equation}\label{partresp}
\chi_{\rm p}=-\frac{\omega_{\rm p}^2}{\omega(\omega+i\nu_p)},
\end{equation}
where $\omega_p=\sqrt{4\pi Q^2 n_p/m}$ is the plasma-particle frequency ($Q$, $n_p$, and $m$ are the particle charge, density, and mass, respectively) and $\nu_p$ is the effective particle-neutral collision frequency (Epstein drag frequency). In the collision-dominated situation we equate the imaginary contributions to the dispersion relation ($\chi_{\rm i}+\chi_{\rm p}=0$) to get in the long-wavelength limit:
\begin{equation}
\omega\approx \omega_{p}\lambda_{i}M\left(\theta_i/\theta_p\right)k=c_{\rm s}k,    \label{sound}
\end{equation}
where $\theta_i=\nu_i/\omega_{{\rm p}i}$ is the ion collisionality parameter  ($\omega_{{\rm p}i}=\sqrt{4\pi e^2 n_i/m_i}$ is the ion-plasma frequency) and $\theta_p=\nu_p/\omega_{p}$ is the particle collisionality parameter. The emerging sound velocity is denoted as $c_{\rm s}$. It is the factor $M(\theta_i/\theta_p)$ that is responsible for the difference between the actual sound velocity $c_{\rm s}$ and the dust-acoustic velocity $c_{\rm DA}=\omega_{p}\lambda_{\rm i}$~\cite{Rao1990} in an idealized isotropic collisionless dusty plasma, with screening dominated by the ion component. For further details see Refs.~\cite{Khrapak2003,Khrapak2020}.

Equation (\ref{sound}) strictly applies when both the ion and the particle components are collision dominated. For ions this requires $M\nu_i/kv_{{\rm T}i}\gg 1$, which is very well satisfied in the parameter regime investigated. For the particle component the wave frequency should be lower than  the Epstein drag frequency. The latter can be estimated as $\nu_p\approx 60$ s$^{-1}$ and we see from Fig. \ref{fig:results2} that this condition is not satisfied. Nevertheless, we have verified that in our parameter regime, the long-wavelength limit of Eq.~(\ref{sound}) remains a very reasonable simple approximation of the phase velocity obtained from a direct solution of the dispersion relation $\chi_{\rm i}+\chi_{\rm p}=0$ at finite wavelengths, especially at lower frequencies. We therefore use it below and this allows us to easily identify the main physical factors behind the variation of the sound velocity.

To perform numerical estimates of the dust sound velocity using Eq.~(\ref{sound}) we use the following set of {\it fixed} dusty plasma parameters (in addition to $r_p=2.2$ $\mu$m and $p=23$ Pa specified above). The electron and ion temperatures are taken as $T_e= 2$ eV and $T_i= 0.03$~eV. The ion density is estimated as $n_i\approx 7\times 10^8$ cm$^{-3}$. These are typical parameters for the pressure and voltage used in this experiment, based on computer simulations of the PK-3 Plus discharge~\cite{Thomas2008}. This results in the ion Debye radius $\lambda_i\approx 50$ $\mu$m.  The reduced particle charge is estimated as $z=(|Q|e/r_pT_e)\approx 0.5$, according to existing experimental results on particle charging in gas discharges at moderate neutral gas pressures~\cite{Ratynskaia2004, Khrapak2005, Antonova2019}. Actually, neither the ion density nor the reduced particle charge are strictly constant across the pre-sheath region; we just expect that their variations are of much smaller significance than the variations in the ion thermal Mach number and collisionality  (see e.g., Ref.~\cite{Khrapak2013}). Some dependence of the particle density on the position in the pre-sheath (that is on the magnitude of the temperature gradient applied) is observed in the experiment and is retained in the calculation. The following practical fit is used: $n_p\approx 8\times 10^4-3\times 10^3 \, \nabla T$, where $n_p$ is expressed in cm$^{-3}$ and $\nabla$T is expressed in K/cm (more on this in the supplementary material).

To calculate the dependence of the sound velocity on the applied temperature gradient we use the following scheme. We assume that the particle's position in the pre-sheath is mainly determined by the competition between the forces of gravity, electrical force and the thermophoretic force. The electric field at the particle position can be estimated from
\begin{equation}
E(\nabla T) = \frac{mg-F_{\rm th}(\nabla T)}{|Q|},
\end{equation}
where $g\approx 980$ cm/s$^2$ is the acceleration due to gravity. The thermal Mach number is then estimated from the modified Frost formula~\cite{Khrapak2019}
\begin{displaymath}
M=A\left[1+\left(B\frac{E}{N}\right)^C\right]^{-1/2C}\frac{E}{N},
\end{displaymath}
where $N$ is the neutral gas number density and $E/N$ is measured in Townsend units (1 Td = $10^{-17}$ V cm$^2$). For argon gas the parameters are $A=0.0168$ Td$^{-1}$, $B=0.0070$ Td$^{-1}$, and $C=1.238$~\cite{Khrapak2019}. The effective ion-neutral collision frequency is estimated from a practical expression for noble ionized gases suggested in Ref.~\cite{Khrapak2013}
\begin{displaymath}
\nu_i=\nu_0(\gamma M+\sqrt{\gamma^2M^2+1}),
\end{displaymath}
where $\gamma = 0.226$ and $\nu_0=8.22\times 10^{-19} N v_{{\rm T}i}^2$ (in CGS units) for argon gas. The resulting quantities are substituted in Eq.~(\ref{sound}) and the sound velocities are evaluated. The results are shown in Fig.~\ref{fig:results1} (in black). The theoretical results are in reasonable semi-quantitative agreement with the experimental observations.

For reference, we quote the plasma parameters evaluated at zero temperature gradient ($\nabla T = 0$) when the particles are closer to the electrode (deep in the pre-sheath). In this special case we have $\omega_p\approx 90$ s$^{-1}$, $\theta_i\approx 1.2$, $\theta_p\approx 0.6$, and $M\approx 4.5$. The resulting sound velocity $c_{\rm s}\approx 3.8$ cm/s is almost a factor of magnitude higher than the dust-acoustic velocity $c_{\rm DA}\approx 0.4$ cm/s, due to the effect of drifting collisional ions.

\begin{figure}
	\includegraphics[width=0.45\textwidth]{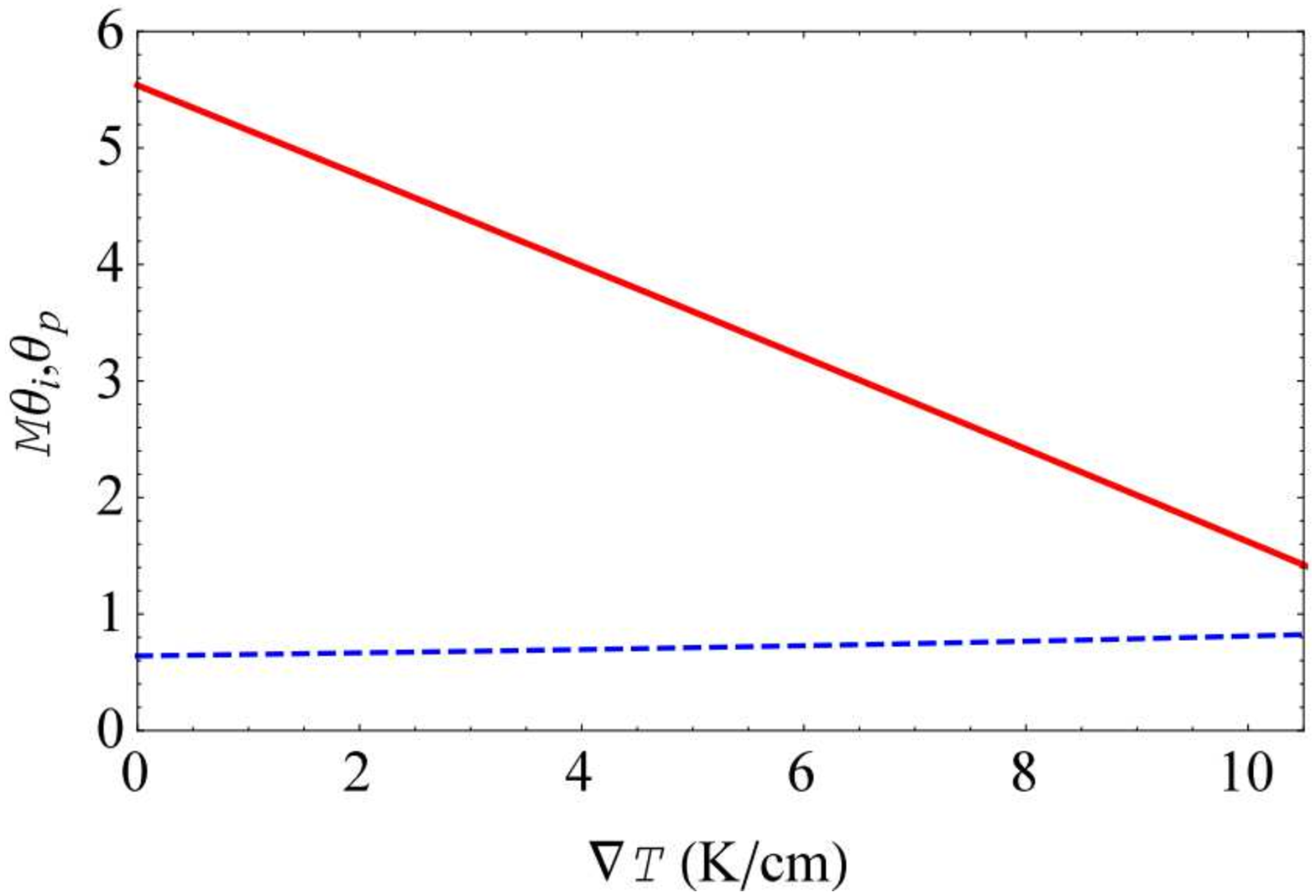}
	\caption{\label{dT} The ion-dependent factor $M\theta_{i}$ (red solid line) and the particle-dependent factor $\theta_p$ (blue dashed line) vs. the temperature gradient $\nabla T$.
    }
\end{figure}

It is easy to identify the main factors that lead to the sound velocity decline when the temperature gradient is applied. With increasing $\nabla T$ the thermophoretic force also increases and hence the electric field needed for particles levitation decreases. This has two consequences: The thermal Mach number $M$ decreases and the effective ion-neutral collisional frequency, roughly proportional to $M$ in the suprathermal ion drift regime, also decreases. The particle-dependent factor $\omega_p/\theta_p\propto \omega_p^2$ in Eq.~(\ref{sound}) also decreases, because the particle density at the position of the cloud decreases with applied temperature gradient. The latter effect is, however, less important. This is illustrated in Fig.~\ref{dT}, which shows the dependence of the ion-dependent and particle-dependent factors $M\theta_i$ and $\theta_p$, respectively, on the temperature gradient. We therefore conclude that the main effect on the DDW velocity is associated with the ion drift velocity variation at the position of the particle cloud (see Fig. \ref{fig:results1}).

\section{\label{sec:conclusion}Conclusion}

We investigated the spatial distribution of dust density wave (DDW) properties in an unmagnetized complex plasma in the ground-based laboratory setup of PK-3 Plus. DDWs were self-excited in the direction of ion drift, towards the bottom electrode. We then heated the bottom electrode to lift the microparticle cloud away from the sheath into the bulk plasma and measured the wave properties as a function of cloud position. The DDW speed and frequency decreased as the cloud moved away from the sheath and into the bulk plasma. We compared the measured wave velocities with the theoretically predicted ones and conclude that the decrease in the DDW velocity might be associated with the change in ion drift velocity with height.

Since an inhomogeneous stationary state of plasmas is closer to physical reality, our findings might be relevant  for the wave studies in  many complex /dusty plasmas, including astrophysical  and fusion applications.   For example, in  space dusty plasmas of  planetary rings,  the magnetized electrons and ions co-rotate with the planet, while the ``heavy” grains are not magnetized and move around the planet with Keplerian velocities. Therefore, the plasma exhibits  an azimuthal drift relative to the ring particles. The drift  velocity depends on the radial distance to the synchronous orbit and varies in a wide range from zero at the synchronous orbit to significant values of the order of the ion thermal  velocity at the boundary of the main rings. Such a radial dependence of the  plasma drift velocity  causes the existence of spatial zones where the dust density waves can be excited and regions where the  DDWs are damped \cite{Yaroshenko2006}.

The results presented in this paper could thus give a new insight into the ion-streaming instability  in the dusty plasma of the planetary rings  and  quantify  the effect in terms of  variations of the ion Mach number.

\begin{acknowledgments}
We acknowledge funding of this work in the framework of the Nachwuchsgruppenprogramm im DLR-Gesch\"aftsbereich Raumfahrtforschung und -technologie. Theoretical work was supported by the Ministry of Science and Higher Education of the Russian Federation
(Agreement with Joint Institute for High Temperatures RAS No 075-15-2020-785 dated September 23, 2020).
We thank Peter Huber and Daniel Mohr for offering IT help during the course of this project; Tanja Hagl and Claudius George for experimental support; and Hubertus Thomas and Alexei Ivlev for their input and support. We would also like to thank Matthias Sperl for careful reading of the manuscript and useful suggestions.
\end{acknowledgments}

%


\end{document}


\title{Supplementary Material} \maketitle
\section{Periodgrams or space-time plots for all temperature gradients} Here, we present the periodgrams at different temperature gradients only for a selected duration of time. The actual time sequences are 26 times longer than presented in each of these periodgrams. In general, we observe that the waves become weaker with increasing temperature gradient as the microparticle cloud moves away from the sheath region and into the bulk region. \\
At some temperature gradients in the images shown below, for example at $\Delta T= 10 K$, we see no wavecrests in the chosen time window. However at a higher temperature gradient, for example at $\Delta T = 12.5 K$, we see the wave crests again. Here, we would like to stress that this is not due to any physical reasons. It is simply because the time sequence shown here does not include any wave crests at $\Delta T = 10 K$. However, if we look at a later time sequence instance at $\Delta T = 10 K$, as shown in Fig. \ref{fig:10K_new}, we can see wave crests at this temperature gradient as well. We must remember that these waves become very weak at higher temperature differences as the dust cloud moves away from the sheath region.

\begin{figure*}[h!]
  \includegraphics[width=0.9\textwidth]{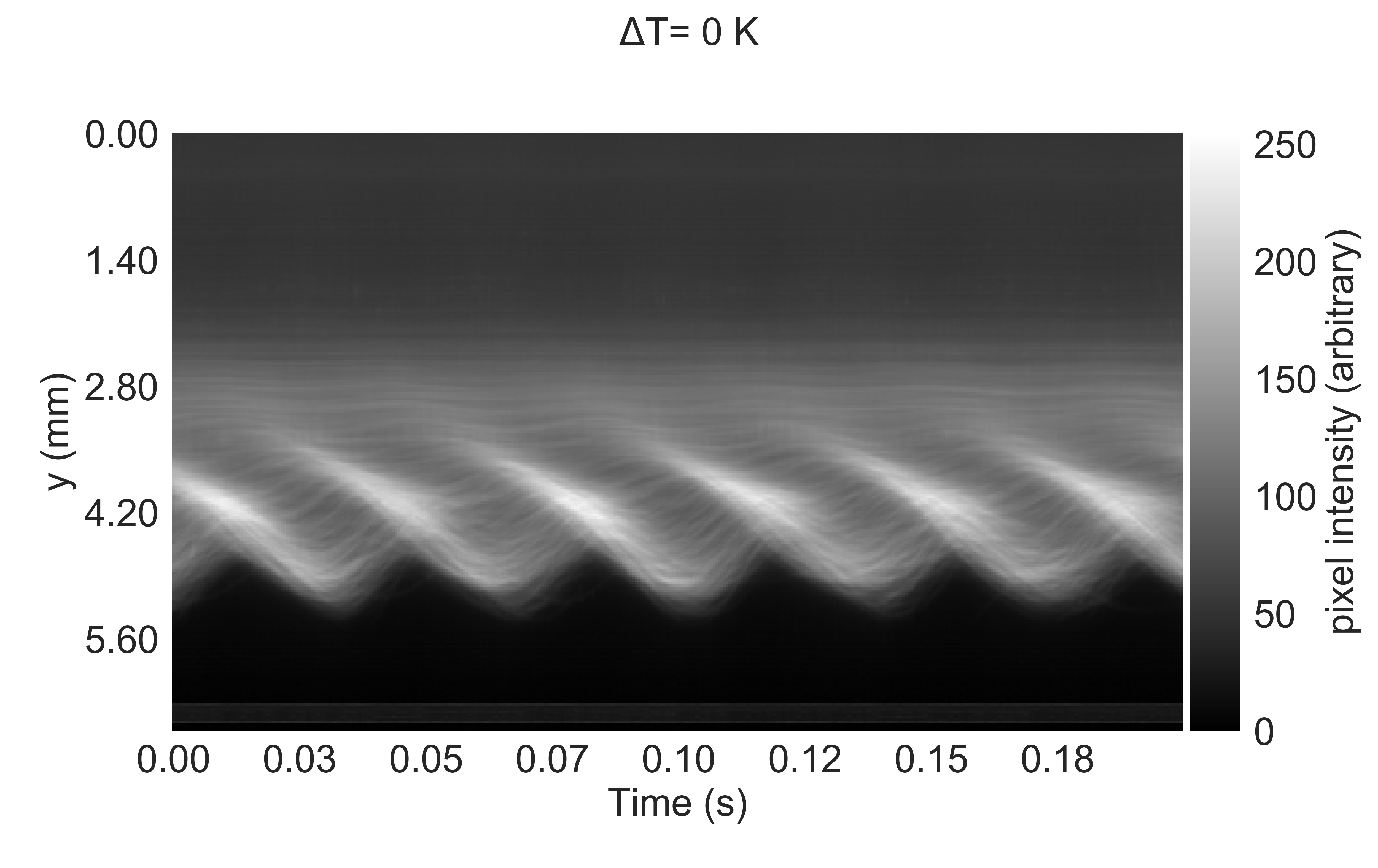}
  \caption{Periodgram or space-time plot at $\Delta T = 0 K$ showing only a small time sequence instance.}
  \label{fig:pdg_0}
\end{figure*}

\begin{figure*}
  \includegraphics[width=0.9\textwidth]{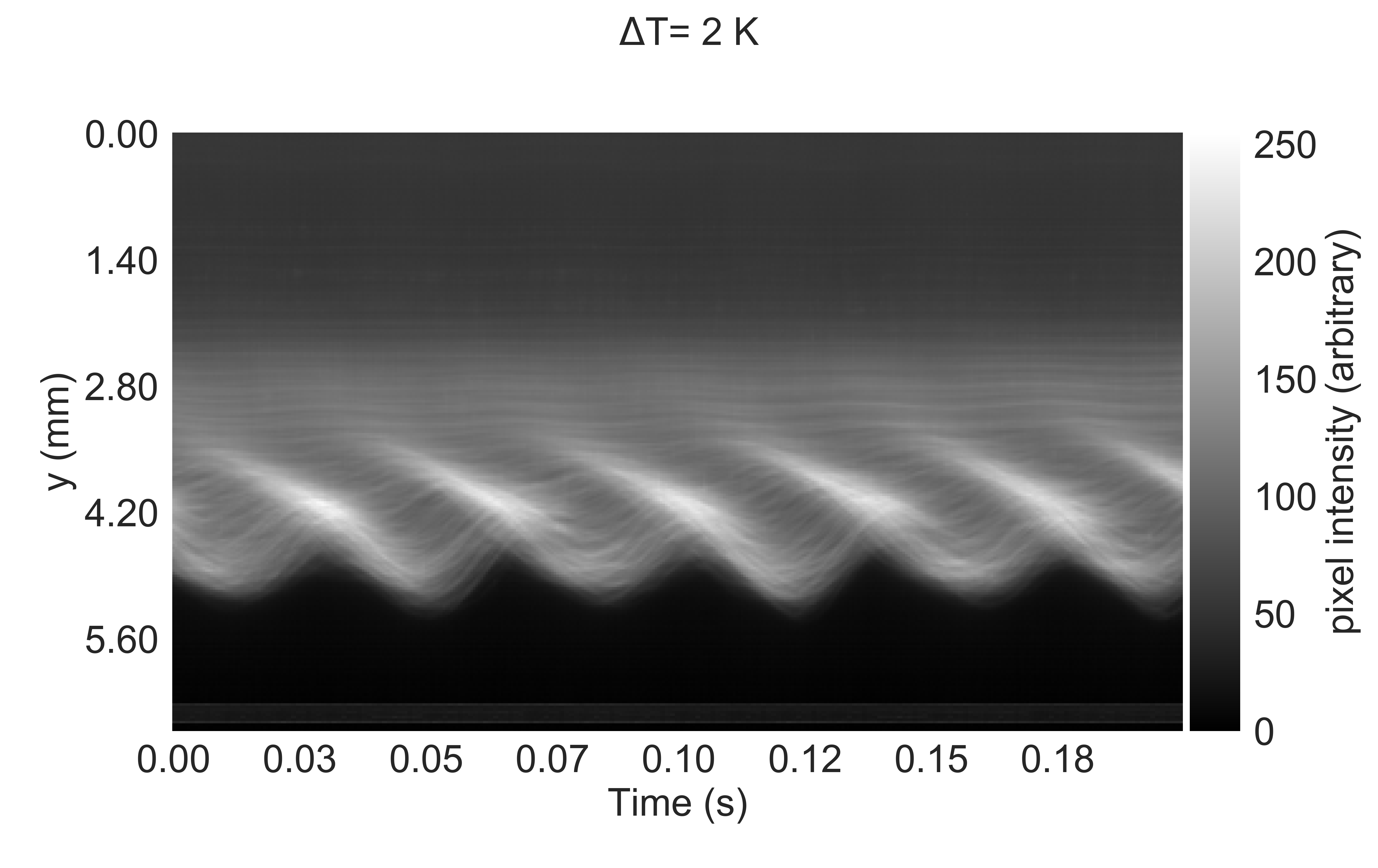}
  \caption{Periodgram or space-time plot at $\Delta T = 2 K$ showing only a small time sequence instance.}
  \label{fig:pdg_2}
\end{figure*}

\begin{figure*}
  \includegraphics[width=0.9\textwidth]{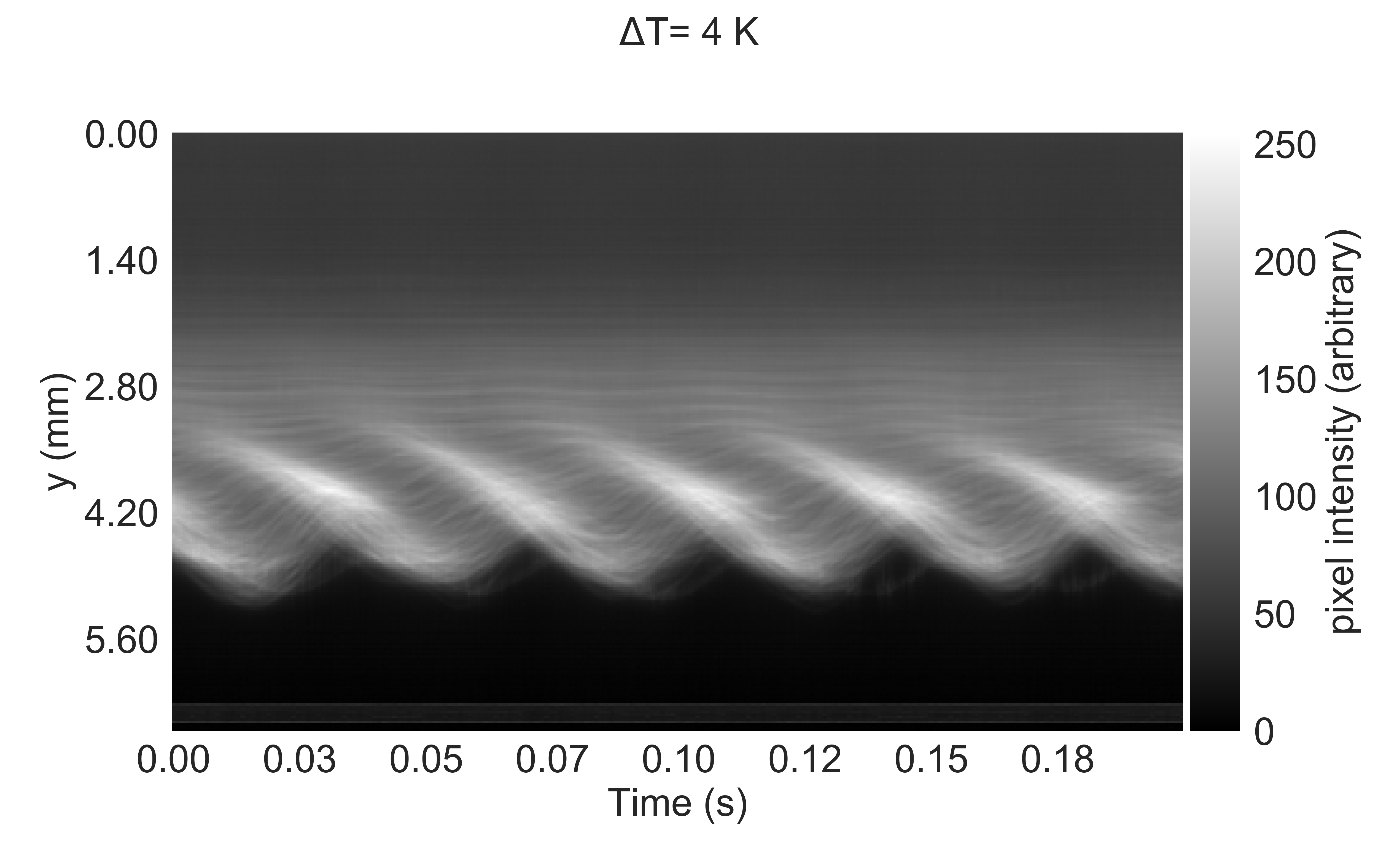}
  \caption{Periodgram or space-time plot at $\Delta T = 4 K$ showing only a small time sequence instance.}
  \label{fig:pdg_4}
\end{figure*}

\begin{figure*}
  \includegraphics[width=0.9\textwidth]{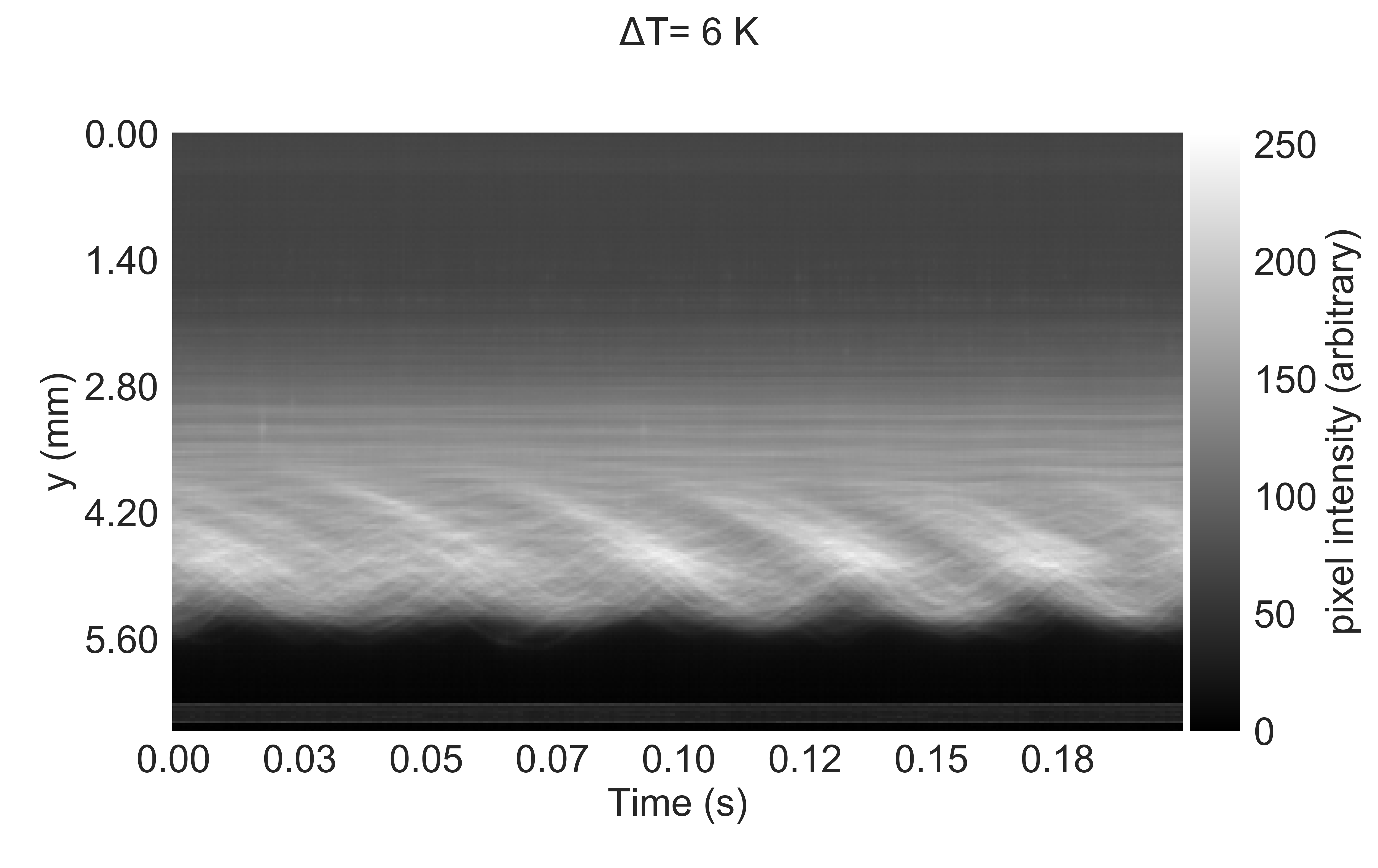}
  \caption{Periodgram or space-time plot at $\Delta T = 6 K$ showing only a small time sequence instance.}
  \label{fig:pdg_6}
\end{figure*}

\begin{figure*}
  \includegraphics[width=0.9\textwidth]{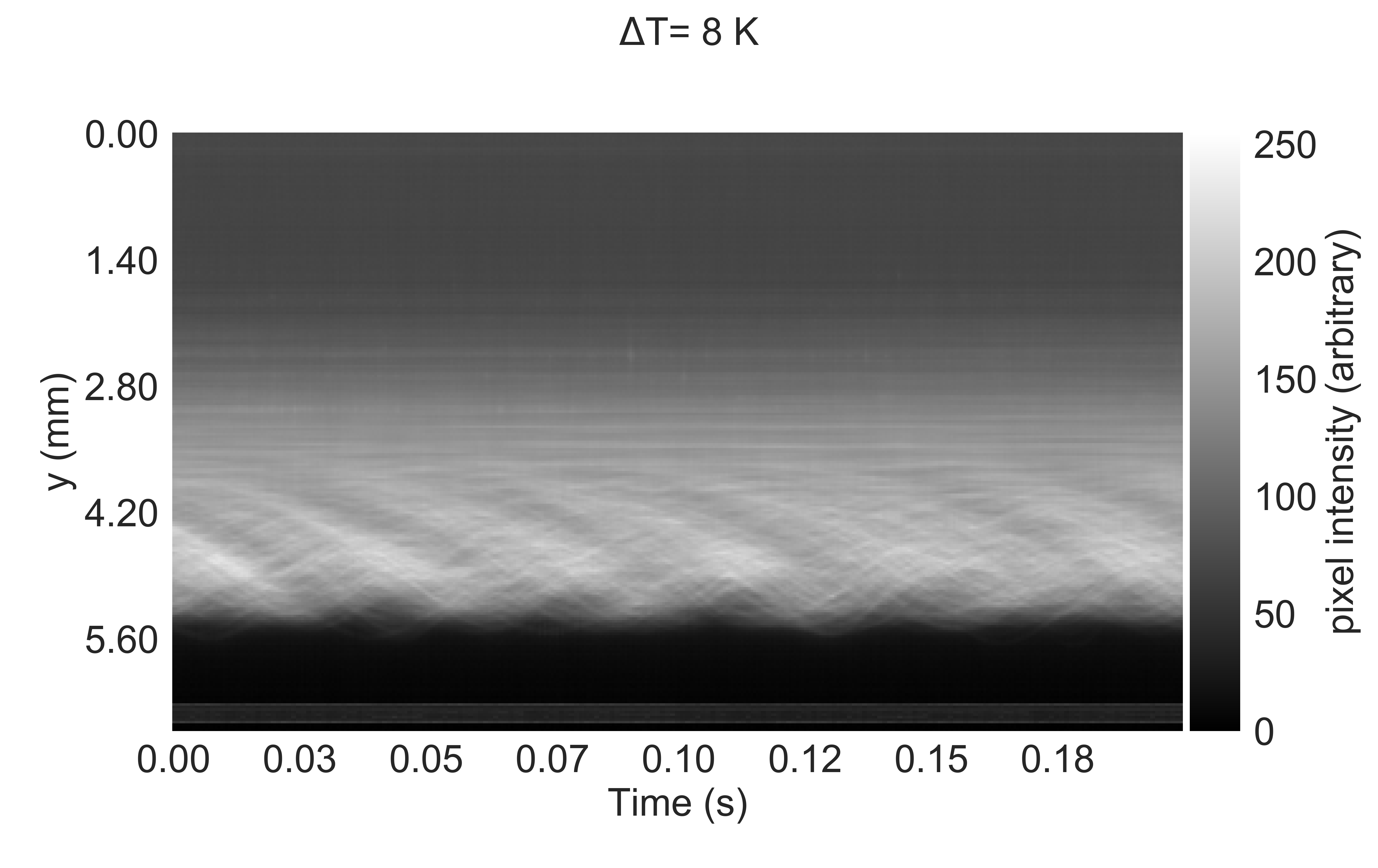}
  \caption{Periodgram or space-time plot at $\Delta T = 8 K$ showing only a small time sequence instance.}
  \label{fig:pdg_8}
\end{figure*}

\begin{figure*}
  \includegraphics[width=0.9\textwidth]{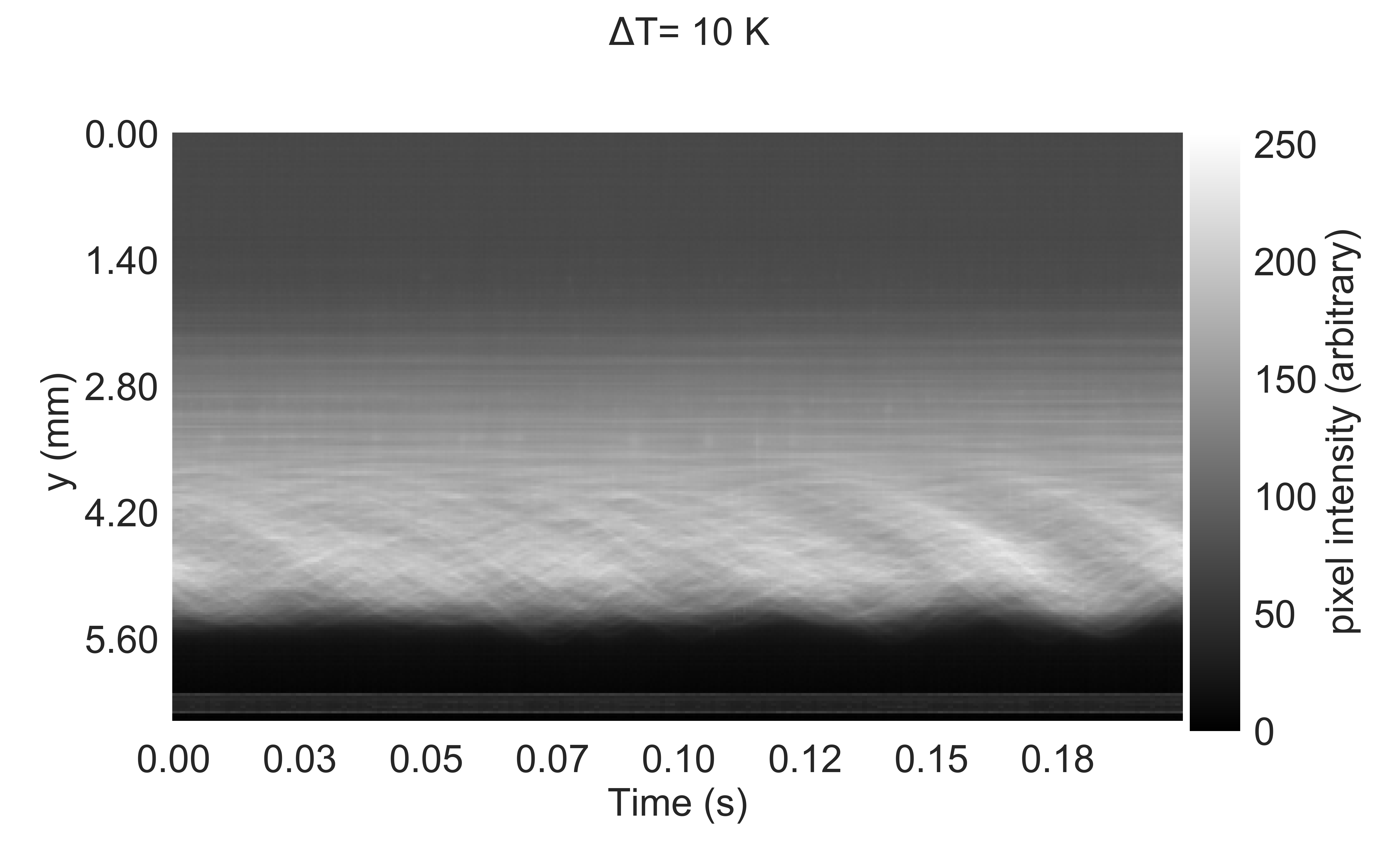}
  \caption{Periodgram or space-time plot at $\Delta T = 10 K$ showing only a small time sequence instance.}
  \label{fig:pdg_10}
\end{figure*}

\begin{figure*}
  \includegraphics[width=0.9\textwidth]{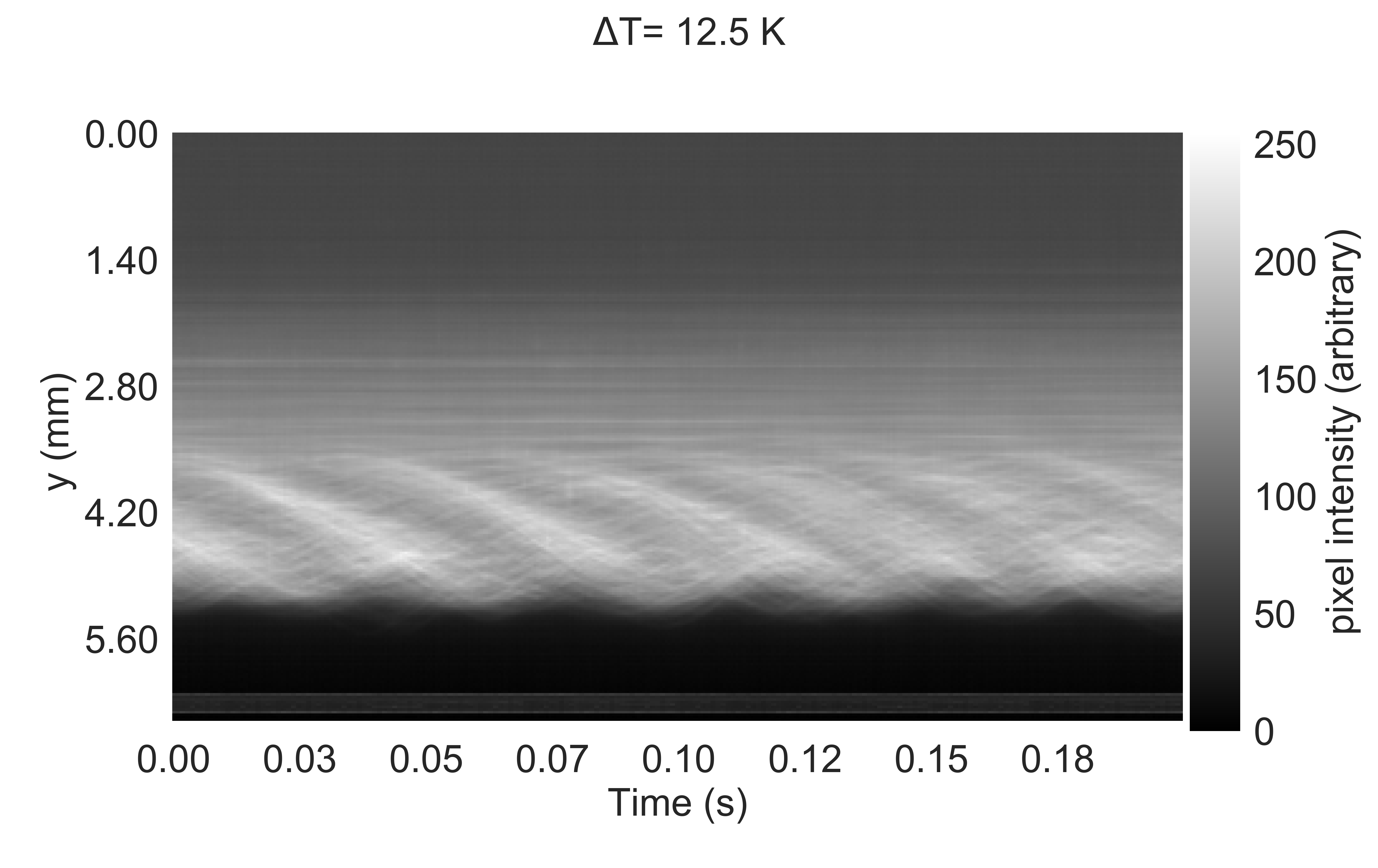}
  \caption{Periodgram or space-time plot at $\Delta T = 12.5 K$ showing only a small time sequence instance.}
  \label{fig:pdg_12.5}
\end{figure*}

\begin{figure*}
  \includegraphics[width=0.9\textwidth]{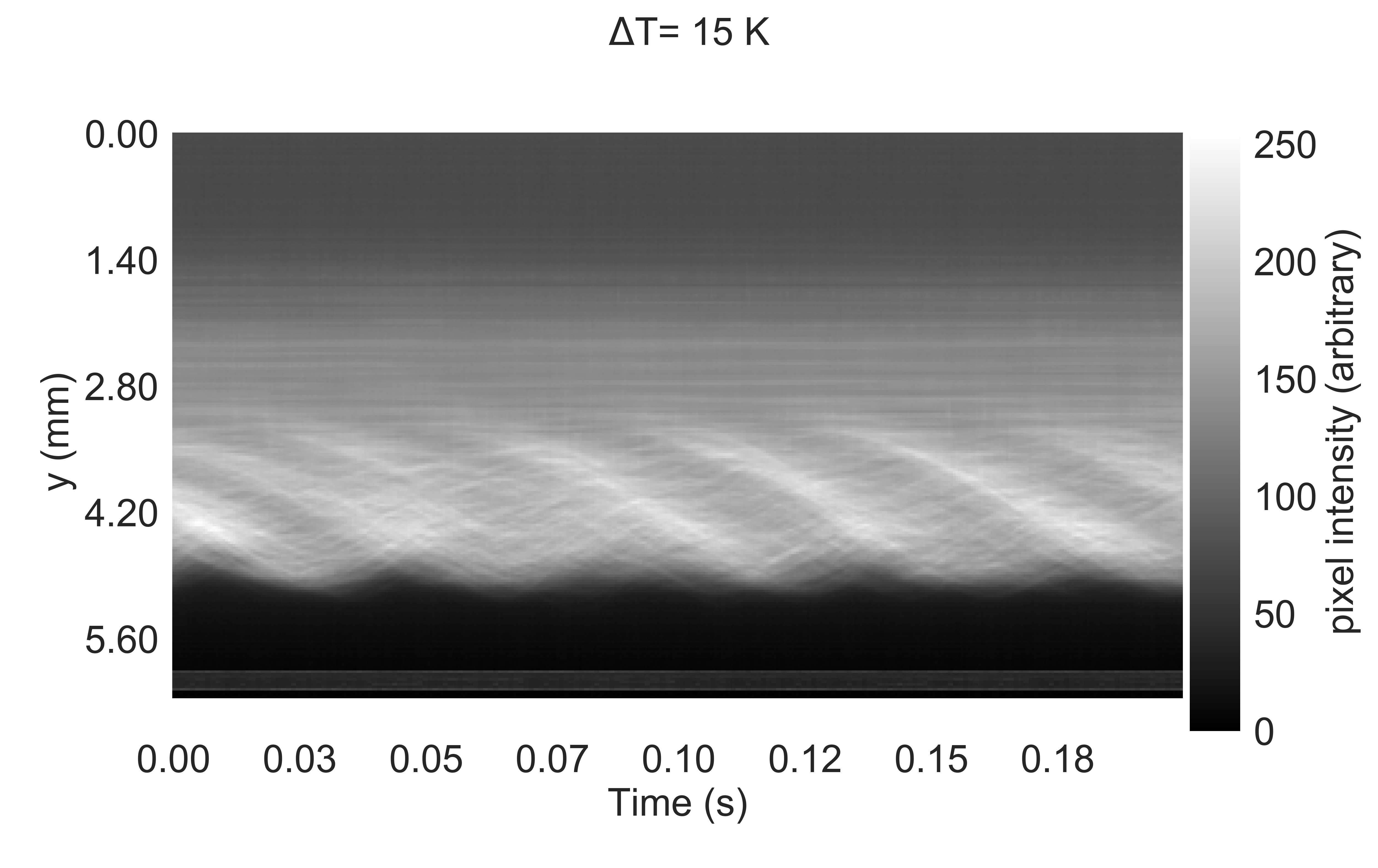}
  \caption{Periodgram or space-time plot at $\Delta T = 15 K$ showing only a small time sequence instance.}
  \label{fig:pdg_15}
\end{figure*}

\begin{figure*}
  \includegraphics[width=0.9\textwidth]{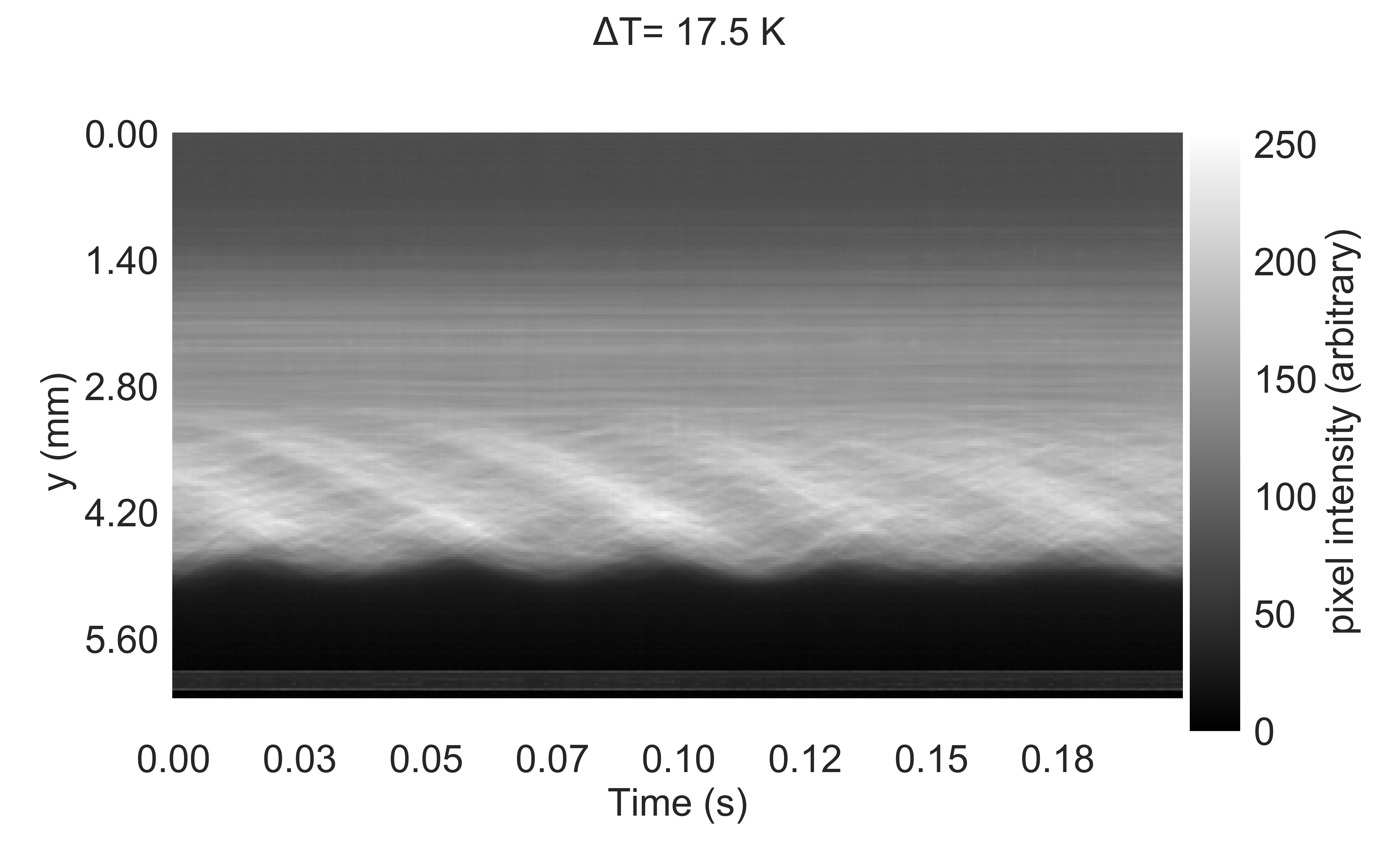}
  \caption{Periodgram or space-time plot at $\Delta T = 17.5 K$ showing only a small time sequence instance.}
  \label{fig:pdg_17.5}
\end{figure*}

\begin{figure*}
  \includegraphics[width=0.9\textwidth]{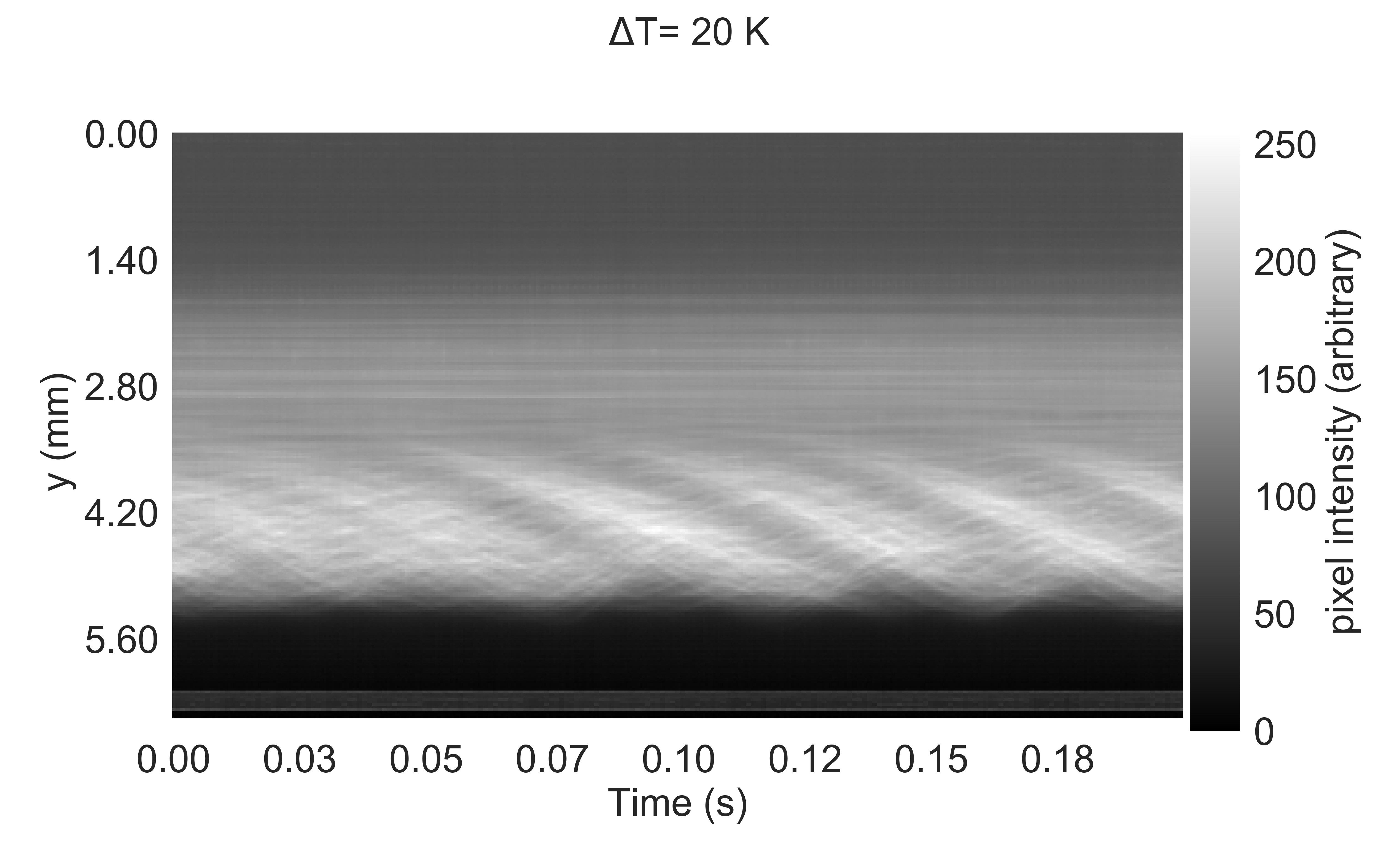}
  \caption{Periodgram or space-time plot at $\Delta T = 20 K$ showing only a small time sequence instance.}
  \label{fig:pdg_20}
\end{figure*}

\begin{figure*}
  \includegraphics[width=0.9\textwidth]{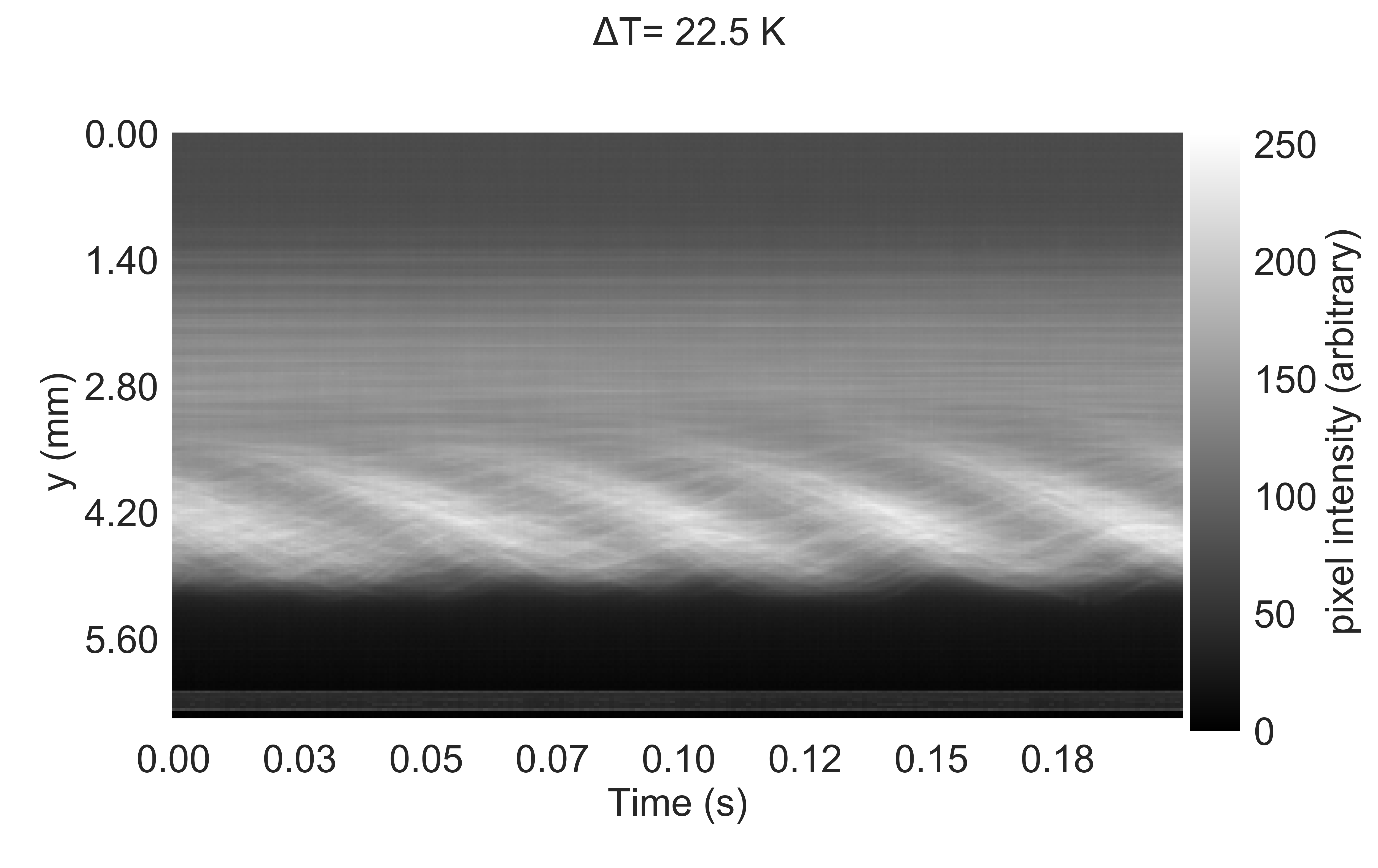}
  \caption{Periodgram or space-time plot at $\Delta T = 22.5 K$ showing only a small time sequence instance.}
  \label{fig:pdg_22.5}
\end{figure*}

\begin{figure*}
  \includegraphics[width=0.9\textwidth]{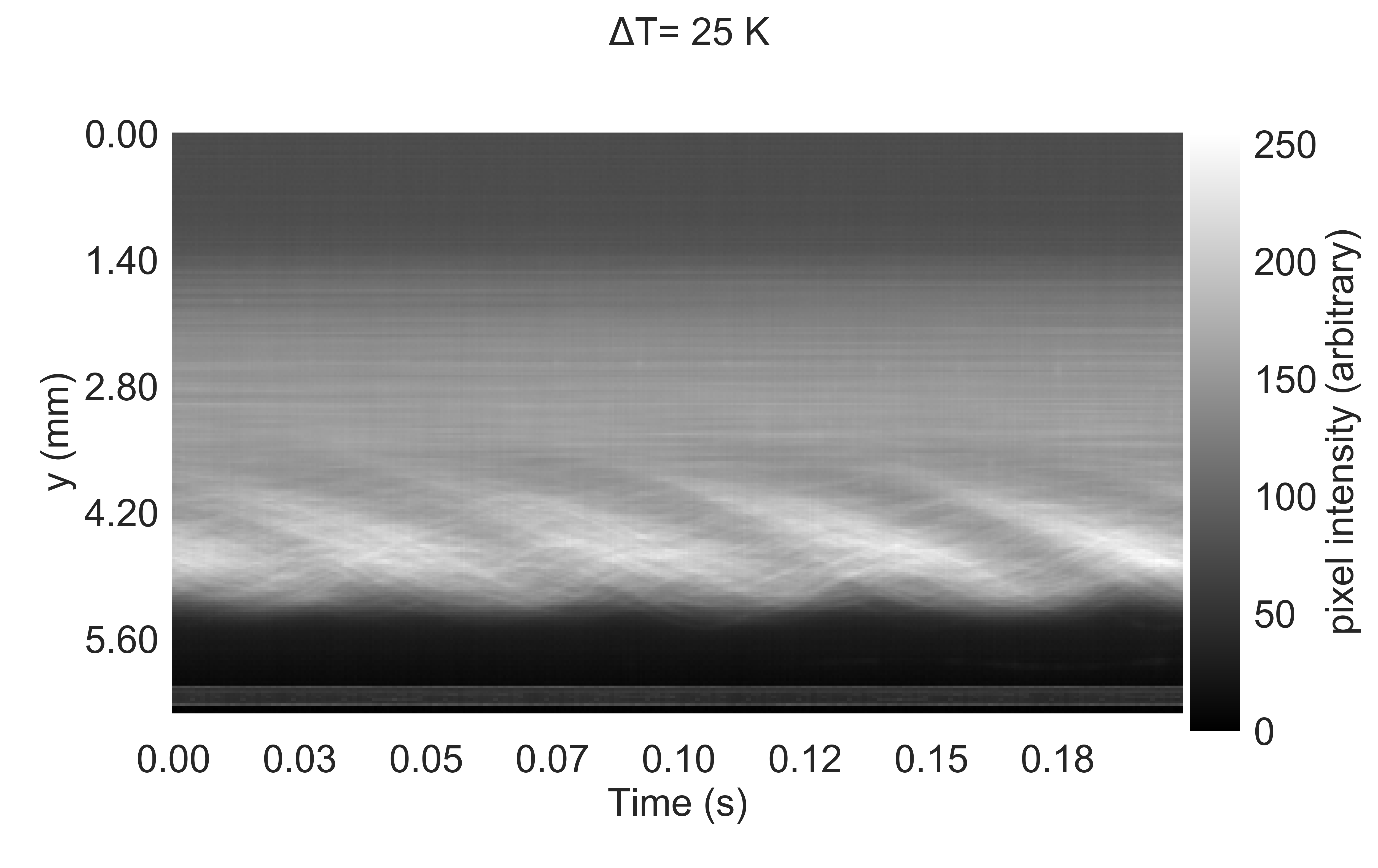}
  \caption{Periodgram or space-time plot at $\Delta T = 25 K$ showing only a small time sequence instance.}
  \label{fig:pdg_25}
\end{figure*}

\begin{figure*}
  \includegraphics[width=0.9\textwidth]{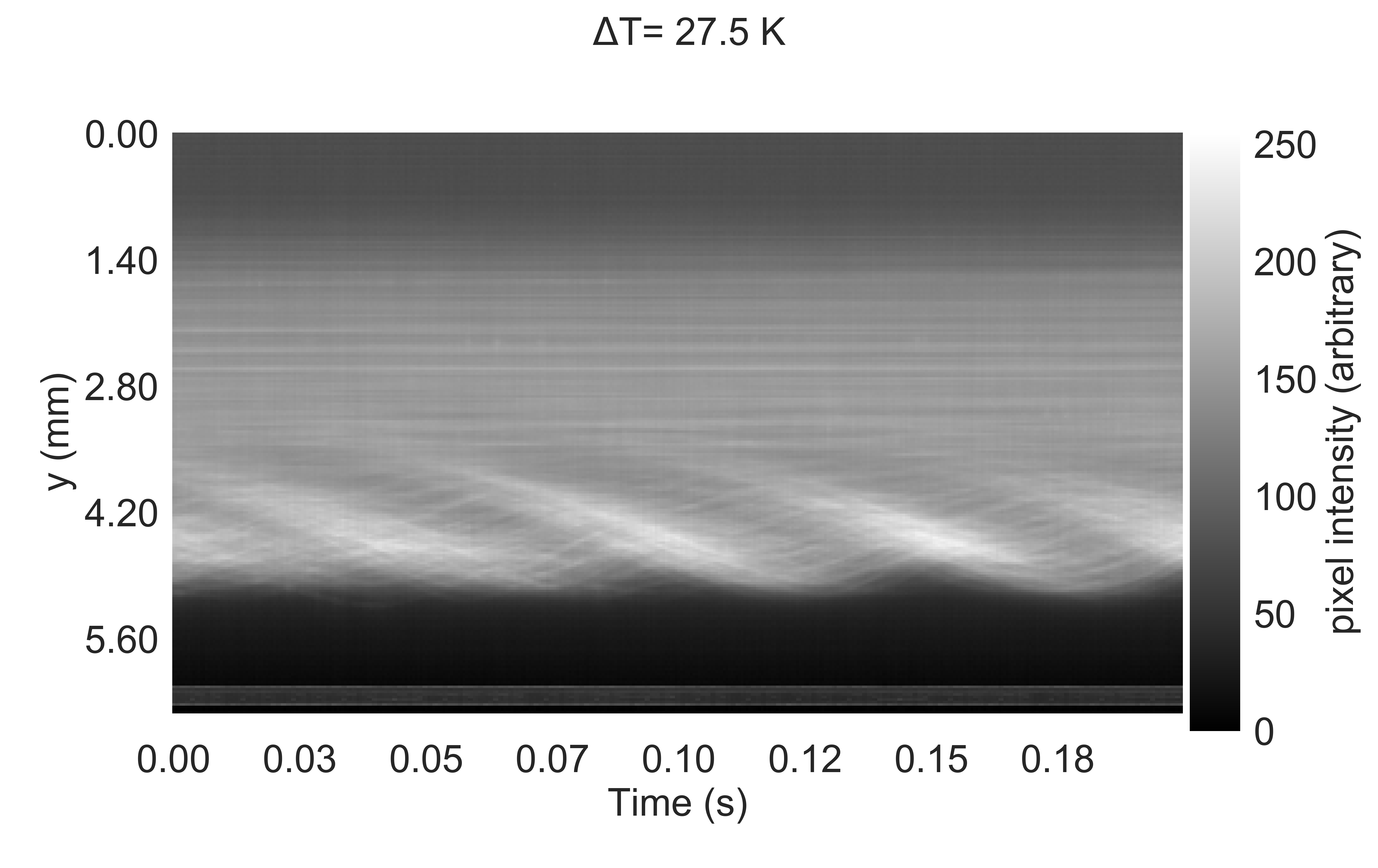}
  \caption{Periodgram or space-time plot at $\Delta T = 27.5 K$ showing only a small time sequence instance.}
  \label{fig:pdg_27.5}
\end{figure*}

\begin{figure*}
  \includegraphics[width=0.9\textwidth]{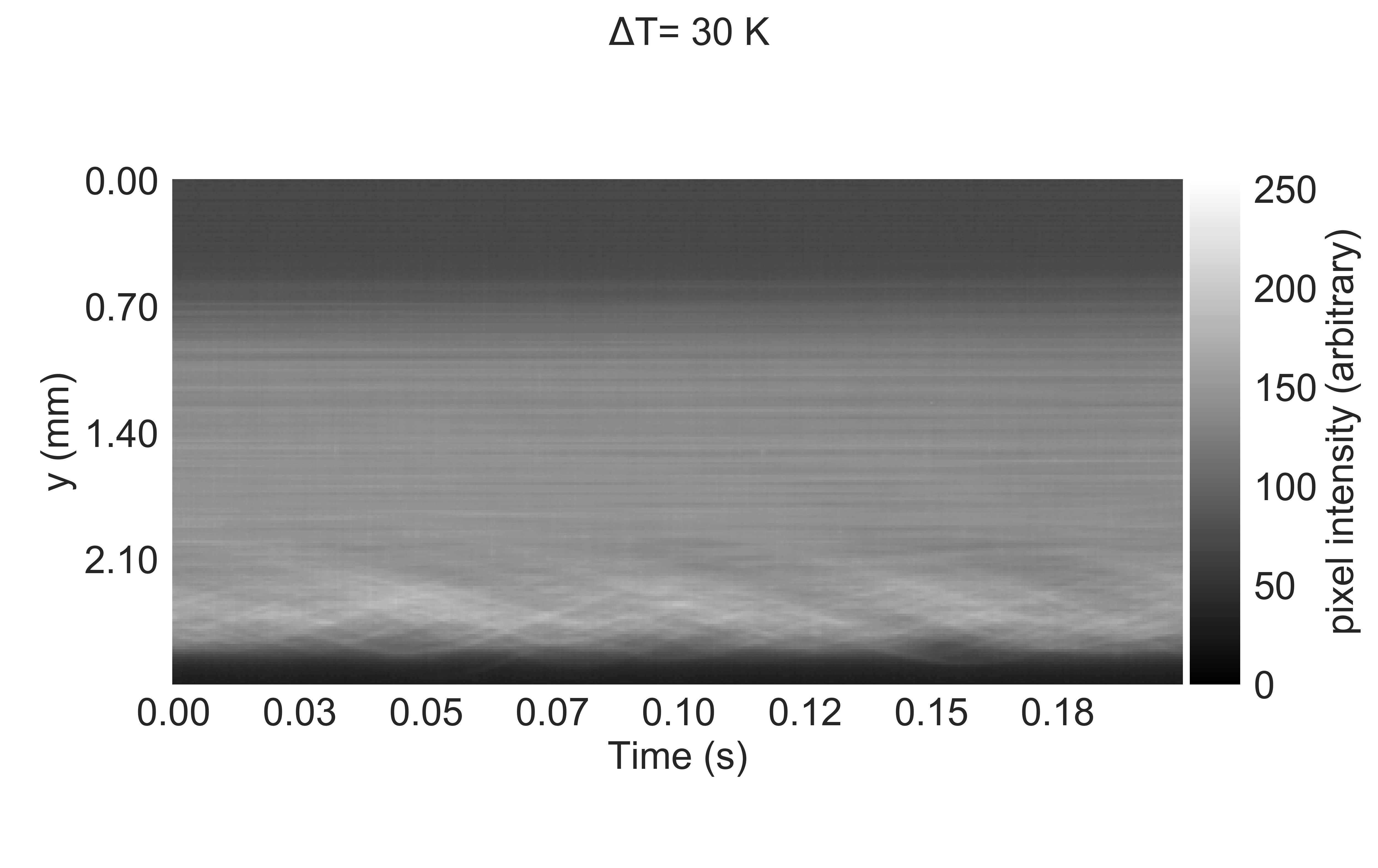}
  \caption{Periodgram or space-time plot at $\Delta T = 30 K$ showing only a small time sequence instance.}
  \label{fig:pdg_30}
\end{figure*}

\begin{figure*}
  \includegraphics[width=0.9\textwidth]{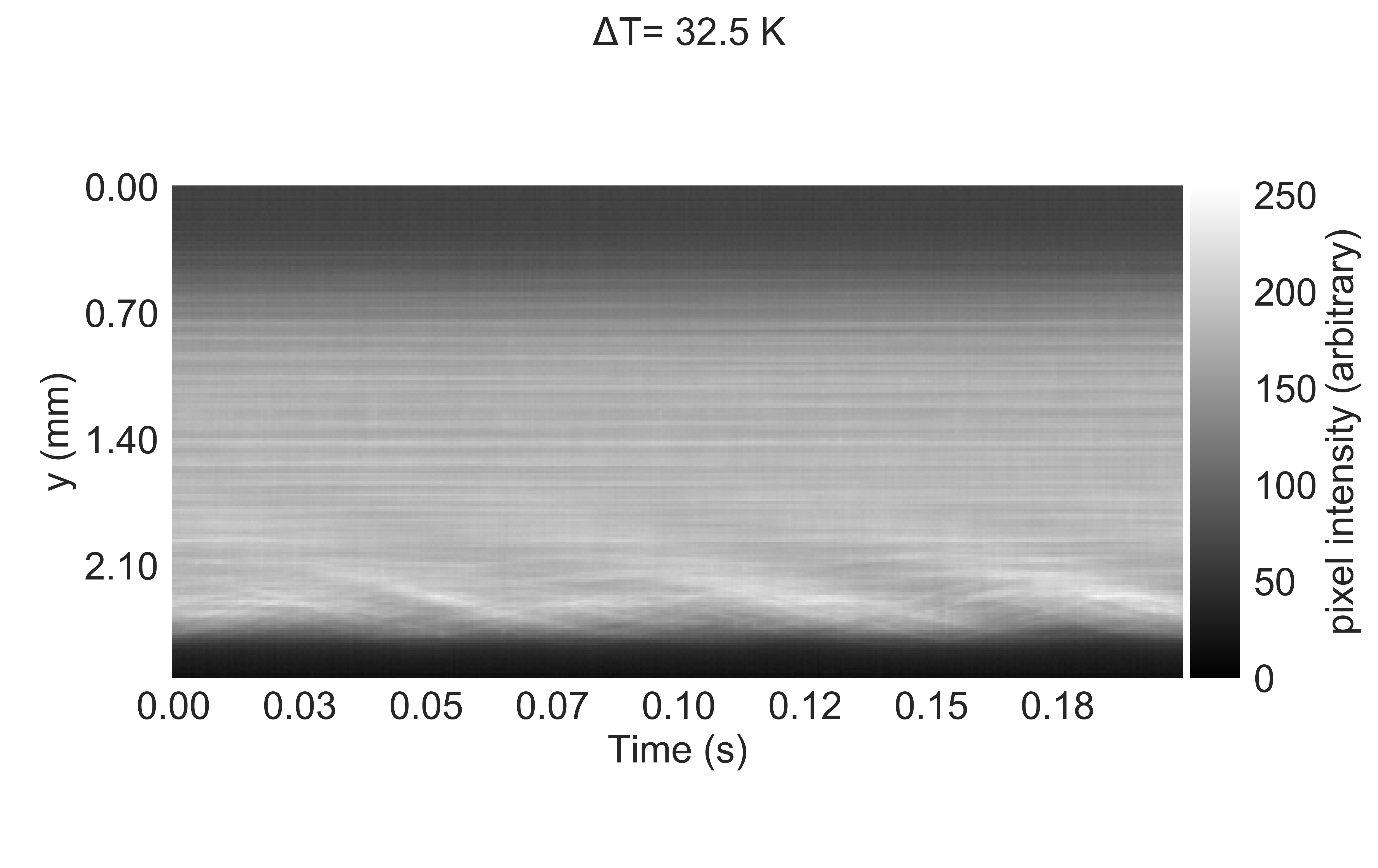}
  \caption{Periodgram or space-time plot at $\Delta T = 22.5 K$ showing only a small time sequence instance.}
  \label{fig:pdg_32.5}
\end{figure*}

\begin{figure*}
  \includegraphics[width=0.9\textwidth]{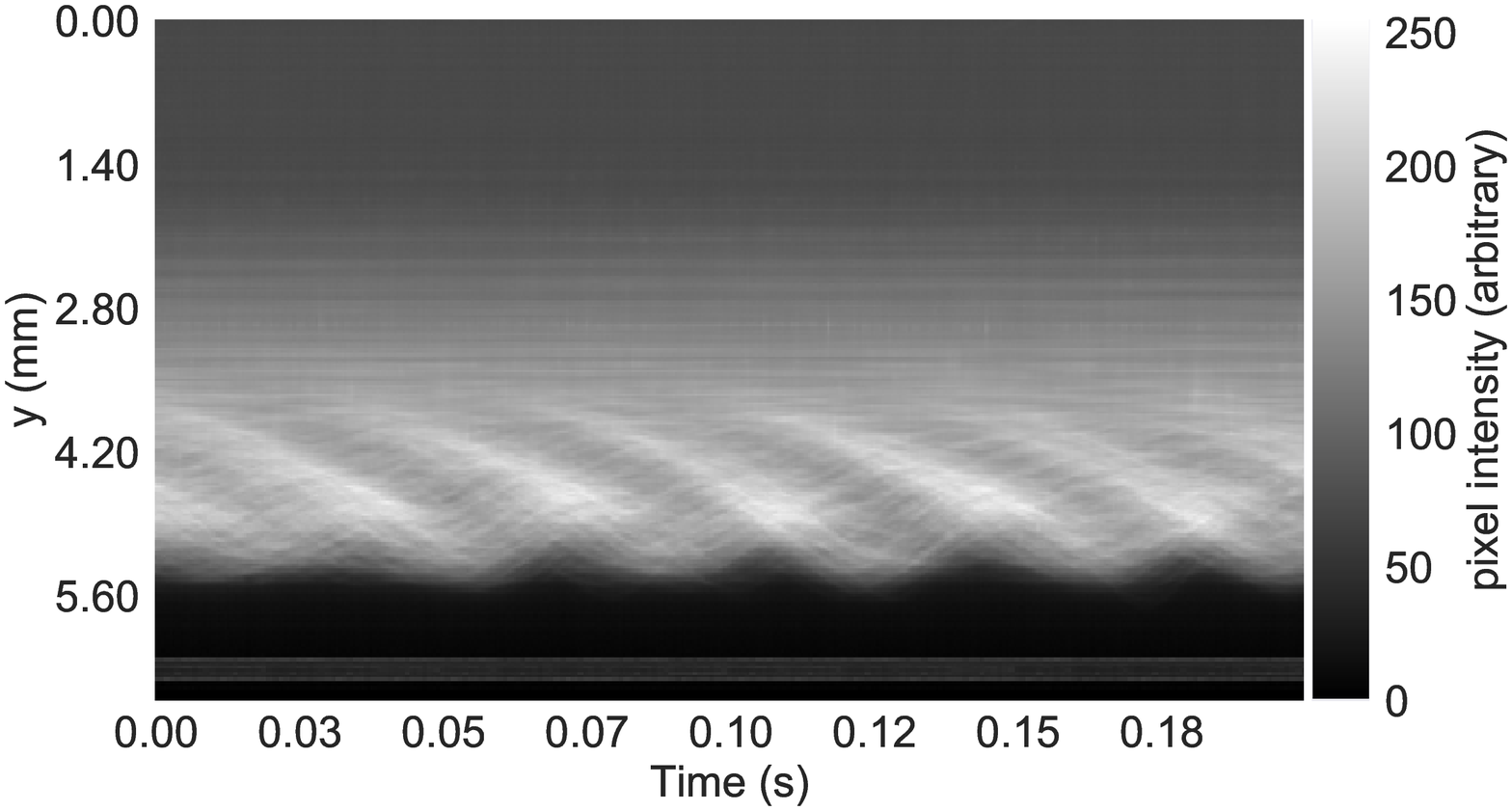}
  \caption{Periodgram or space-time plot at $\Delta T = 10 K$ showing only a small time sequence instance, which is different from the time sequence instance shown in Fig. \ref{fig:pdg_10}.}\label{fig:10K_new}
\end{figure*}

\clearpage

\newpage\section{Measuring wave frequency using template matching}
\begin{figure*}[hb!]
  \includegraphics[width=0.9\textwidth]{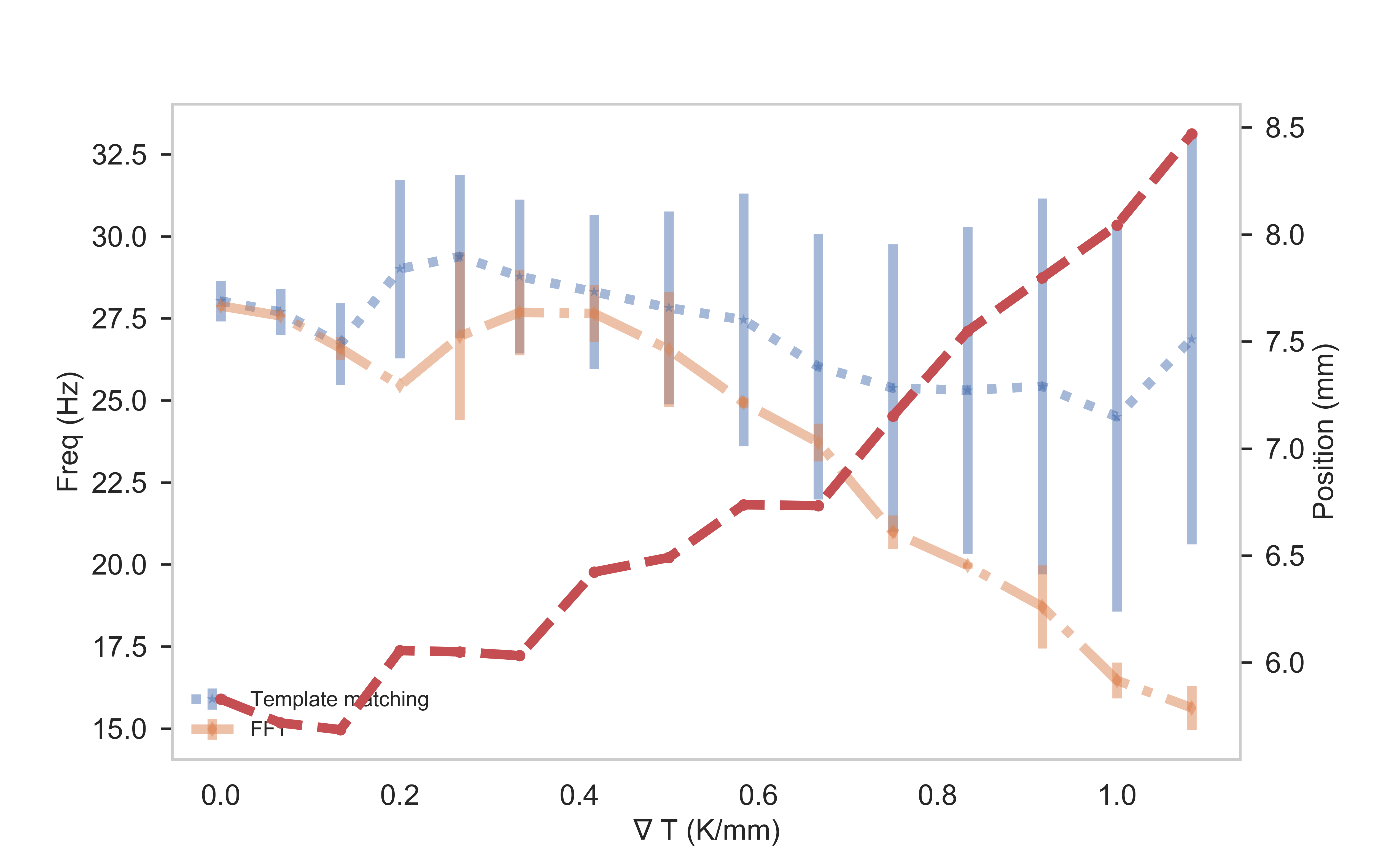}
  \caption{Average wave frequency versus temperature gradient. The line in blue shows the frequency found using the method of template matching and the line in orange shows the frequency found by making a Fast Fourier Transform (FFT) of the change in average particle velocity with time (refer Fig. 5 in the manuscript). The vertical lines are the standard deviations. It is evident from this plot that the standard deviation is extremely high for the template matching method, especially at high temperature gradients. This is the reason we choose to report our results only using the method of fourier transform of tracked particle velocities in time.}\label{fig:std_freq}
\end{figure*}

As mentioned in the manuscript in Section IIB and shown here in Fig. \ref{fig:std_freq}, the results for frequency calculations using template matching gives very high standard deviation. The reason for this is that the wave crest is not always clear, for instance at $\Delta T = 10 K$ in Fig. \ref{fig:pdg_10}, where we observe only 2 wave crests in the selected time region. In such cases, the template matching method is not able to find two consecutive crests, which leads to very high error in calculation of average DDW frequency.

\clearpage

\newpage\section{Calculation of practical fit for particle number density v/s temperature gradient}
\begin{wrapfigure}{l}{0.5\textwidth}
  \includegraphics[width=0.5\textwidth]{number_density2.png}
  \caption{Variation of particle number density with temperature difference calculated using raw images from the experiment.}\label{fig:num_dense}
\end{wrapfigure}
The calculation of the change in dust number density with temperature gradient was done using experimental images by calculating the pair correlation function of located particles in the raw image at each temperature difference selected for the experiment. Fig. \ref{fig:num_dense} shows the change in particle number density with temperature difference, calculated using experimental images. After performing a fit on this data, the following correlation was found between the particle number density and temperature gradient: $n_p\approx 8\times 10^4-3\times 10^3 \, \nabla T$, where $n_p$ is expressed in cm$^{-3}$ and $\nabla$T is expressed in K/cm. Hence, this can be used as a practical fit for the very basic theoretical estimates performed in the paper in Section IIIB.